\documentstyle[sprocl,epsfig]{article}

\def\be{\begin{equation}}
\def\ee{\end{equation}}
\def\bea{\begin{eqnarray}}
\def\eea{\end{eqnarray}}

\begin{document}

\title{EFFECTIVE INTERACTIONS ARE EFFECTIVE INTERACTIONS}

\author{BARRY R. HOLSTEIN}

\address{Department of Physics, University of Massachusetts\\ 
Amherst, MA 01003, USA\\E-mail: holstein@physics.umass.edu}


\maketitle
\abstracts{The subject of effective interactions is introduced and
applications in both quantum mechanics and quantum field theory are
presented.  In particular the use of chiral perturbation theory as an
effective low energy description of QCD is developed and it is argued
that such methods really are effective in both the meson and baryonic sectors.}

\section{Introduction}
The subject of effective interactions has become widespread in physics and 
the goal of these lectures is to try to present the subject in a somewhat
didactic fashion within the context of low energy quantum 
chromodynamics---QCD.  I have chosen my 
title---"Effective Interactions are Effective Interactions"---in order to 
emphasize the feature that the word "effective" has two quite different 
meanings in English and that {\it both} are relevant to our discussion.  One 
meaning emphasizes that the interaction which is being used is {\it not} 
the real thing but rather tries to mock up the manifestations of the true 
underlying dynamics (in our case QCD).  The other meaning is that the 
interaction is effective---it works!---and I will spend the rest of my time
trying to convince you that both meanings are valid.

\section{What is an Effective Interaction?}

The power of effective field theory is associated with the feature that there
exist many situations in physics involving {\it two scales}, one heavy and one 
light.  Then, provided one is working at energies small compared to
the heavy scale, it is possible to 
fully describe the interactions in terms of an ``effective'' picture,
which is written only in terms of the light degrees of freedom, but which
fully includes the 
influence of the heavy mass scale through its virtual effects.  
A number of very nice review articles on effective field theory can be
found in ref. \cite{eftr}.  
 
In quantum field theory it is easy to represent what is going on.
If $\phi,\,\Phi$ represents the light, heavy field respectively, then the
functional integral which characterizes the full quantized theory is given by
\begin{equation}
W=\int[d\phi][d\Phi]exp\, i\int d^4x{\cal L}(\phi,\Phi)
\end{equation}
Now suppose we integrate out the heavy degrees of freedom----what is left is
a functional integral in terms of a generally non-local "effective" 
interaction which 
characterizes the theory in terms of only the light degree of freedom $\phi$
\begin{equation}
W=N\int[d\phi]\exp i\int d^4x{\cal L}_{eff}(\phi)
\end{equation}
Now while this procedure is formally correct, the physics of what is going on]
is also somewhat nonintuitive, at least to me.  So before proceeding to the
complex subject of application to QCD, it is useful to study this idea 
in the simpler context of ordinary quantum mechanics, in order to get familiar 
with the idea.
Specifically, we examine the question of why the sky is blue, whose 
answer can be
found in an analysis of the scattering of photons from the sun 
by atoms in the atmosphere---Compton scattering\cite{skb}.  First we examine 
the problem using traditional quantum mechanics and, for simplicity, consider 
elastic (Rayleigh) scattering from
single-electron (hydrogen) atoms.  The appropriate Hamiltonian is then
\begin{equation}
H={(\vec{p}-e\vec{A})^2\over 2m}+e\phi
\end{equation}
and the leading---${\cal O}(e^2)$---amplitude for Compton scattering
is found from calculating the diagrams shown in Figure 1, yielding the
familiar Kramers-Heisenberg form
\begin{eqnarray}
{\rm Amp}&=&-{e^2/m\over \sqrt{2\omega_i2\omega_f}}\left[\hat{\epsilon}_i\cdot
\hat{\epsilon}_f^*+{1\over m}\sum_n\left({\hat{\epsilon}_f^*\cdot
<0|\vec{p}e^{-i\vec{q}_f\cdot\vec{r}}|n>
\hat{\epsilon}_i\cdot <n|\vec{p}e^{i\vec{q}_i\cdot\vec{r}}|0>\over 
\omega_i+E_0-E_n}\right.\right.
\nonumber\\
&+&\left.\left.{\hat{\epsilon}_i\cdot <0|\vec{p}e^{i\vec{q}_i\cdot\vec{r}}|n>
\hat{\epsilon}_f^*\cdot <n|\vec{p}e^{-i\vec{q}_f\cdot\vec{r}}|0>\over E_0-\omega_f-E_n}
\right)\right]
\end{eqnarray}
where $|0>$ represents the hydrogen ground state having binding energy
$E_0$. 
\begin{figure}
\begin{center}
\epsfig{file=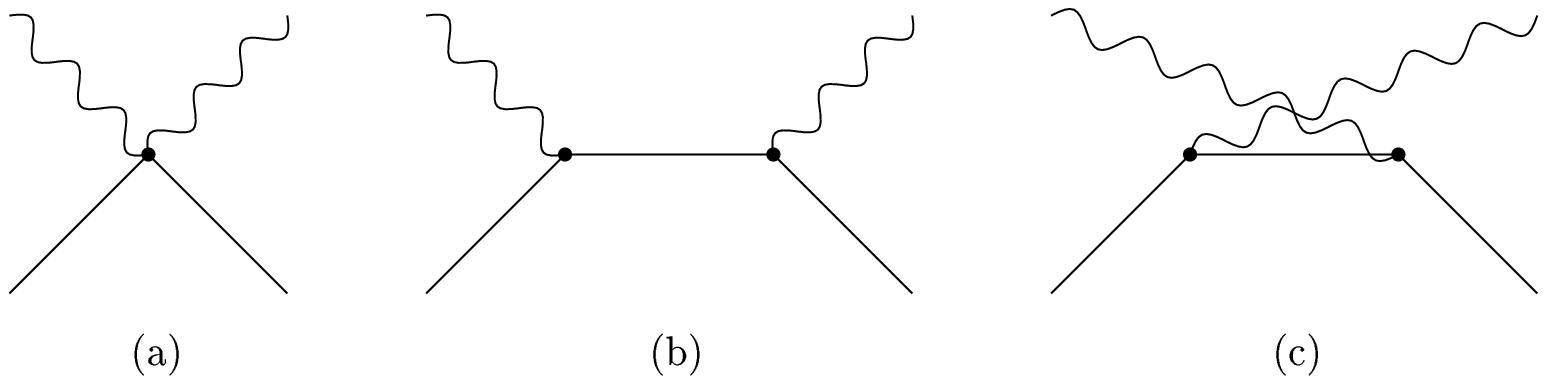,height=4cm,width=12cm}
\caption{Feynman diagrams for nonrelativistic photonl-atom scattering.}
\end{center}
\end{figure}

Here the leading component is the familiar 
$\omega$-independent Thomson amplitude and would appear naively 
to lead to an energy-independent
cross-section.  However, this is {\it not} the case.  Indeed, by
expanding in $\omega$ and using a few quantum mechanical identities
one can show that, 
provided that the energy of the photon is much smaller than a typical 
excitation energy---as is the case for optical photons, the cross section 
can be written as
\begin{eqnarray}
\quad {d\sigma\over d\Omega}&=&
\lambda^2\omega^4|\hat{\epsilon}_{f}^*\cdot
\hat{\epsilon}_{i}|^2\left(1+{\cal O}\left({\omega^2\over (\Delta E)^2}
\right)\right)\label{eq:cc}
\end{eqnarray}
where 
\begin{equation}
\lambda=\alpha_{em}\sum{2|z_{n0}|^2\over E_n-E_0}
\end{equation}
is the atomic electric polarizability,
$\alpha_{em}=e^2/4\pi$ is the fine structure constant, and $\Delta E\sim m
\alpha_{em}^2$ is a typical hydrogen excitation energy.  We note that  
$\alpha_{em}\lambda\sim a_0^2\times {\alpha_{em}\over \Delta E}\sim a_0^3$ 
is of order the atomic volume, as will be exploited below, and that the cross 
section itself has the characteristic $\omega^4$ dependence which leads 
to the blueness of the sky---blue light scatters much
more strongly than red\cite{feyn}.

Now while the above derivation is certainly correct, it requires somewhat detailed and
lengthy quantum mechanical manipulations which obscure the relatively simple physics
involved.  One can avoid these problems by the use of effective field theory 
methods outlined above.  The key point is that of scale.  Since the incident photons
have wavelengths $\lambda\sim 5000$A much larger than the $\sim$ 1A atomic size, then at 
leading order the photon is insensitive to the presence of the atom, since the latter
is electrically neutral.  If $\chi$ represents the wavefunction of the atom then the 
effective leading order Hamiltonian is simply that for the hydrogen atom 
\begin{equation}
H_{eff}^{(0)}=\chi^*\left({\vec{p}^2\over 2m}+e\phi\right)\chi
\end{equation}
and there is {\it no} interaction with the field.  In higher orders, 
there {\it can} exist such atom-field interactions and this is
where the effective Hamiltonian comes in to play.  In order to construct the
effective interaction, we demand certain general principles---the
Hamiltonian must satisfy fundamental symmetry requirements.  In
particular $H_{eff}$ must be gauge invariant, must be a scalar under
rotations, and must be even under both parity and time reversal 
transformations.  Also,
since we are dealing with Compton scattering, $H_{eff}$ must be
quadratic in the vector potential.
Actually, from the requirement of gauge invariance it is
clear that the effective interaction should involve only the electric
and magnetic fields, rather than the vector potential itself---
\begin{equation}
\vec{E}=-\vec{\nabla}\phi-{\partial\over \partial t}\vec{A}, 
\qquad \vec{B}=\vec{\nabla}\times\vec{A}\label{eq:ii}
\end{equation}
since these are invariant under a gauge transformation
\begin{equation}
\phi\rightarrow\phi+{\partial\over \partial t}\Lambda,\qquad \vec{A}
\rightarrow\vec{A}-\vec{\nabla}\Lambda
\end{equation}
while the vector and/or scalar potentials are not.  The lowest order
interaction then can involve only the rotational invariants 
$\vec{E}^2,\vec{B}^2$
and $\vec{E}\cdot\vec{B}$.  However, under spatial
inversion---$\vec{r}\rightarrow -\vec{r}$---electric and magnetic
fields behave oppositely---$\vec{E}\rightarrow -\vec{E}$ while
$\vec{B}\rightarrow\vec{B}$---so that parity invariance rules out any
dependence on $\vec{E}\cdot\vec{B}$.  Likewise under time
reversal invariance $\vec{E}\rightarrow\vec{E},\,\,\vec{B}\rightarrow
-\vec{B}$ so such a term is also T-odd.  
The simplest such effective Hamiltonian must
then have the form
\begin{equation}
H_{eff}^{(1)}=\chi^*\chi[-{1\over 2}c_E\vec{E}^2
-{1\over 2}c_B\vec{B}^2]\label{eq:ll}
\end{equation}
(Forms involving time or spatial derivatives are much smaller.)
We know from electrodynamics that 
${1\over 2}(\vec{E}^2+\vec{B}^2)$
represents the field energy per unit volume, so by dimensional
arguments, in order to represent an
energy in Eq. \ref{eq:ll}, $c_E,c_B$ must have dimensions of volume.
Also, since the photon has such a
long wavelength, there is no penetration of the atom, so  only classical scattering
is allowed.  The relevant scale must then be atomic size so that we can write
\begin{equation}
c_E=k_Ea_0^3,\qquad c_B=k_Ba_0^3
\end{equation}
where we expect $k_E,k_B\sim {\cal O}(1)$.  Finally, since for photons
with polarization $\hat{\epsilon}$ and four-momentum $q_\mu$ we
identify $\vec{A}(x)=\hat{\epsilon}\exp(-iq\cdot x)$
then from Eq. \ref{eq:ii}, $|\vec{E}|\sim \omega$, 
$|\vec{B}|\sim |\vec{k}|=\omega$ and 
\begin{equation}
{d\sigma\over d\Omega}\propto|<f|H_{eff}|i>|^2\sim\omega^4 a_0^6
\end{equation}
as found in the previous section via detailed calculation.  This is a
nice example of the power of simple effective field theory arguments.

\subsection{Euler-Heisenberg Lagrangian}

Now consider a second example---photon-photon scattering.  In this case, since the photon
couples to charge but is itself uncharged, there exists no lowest order 
interaction.  However, the $\gamma\gamma\rightarrow\gamma\gamma$ process can proceed 
via the charged particle box diagram shown in Figure 2.  

\begin{figure}
\begin{center}
\epsfig{file=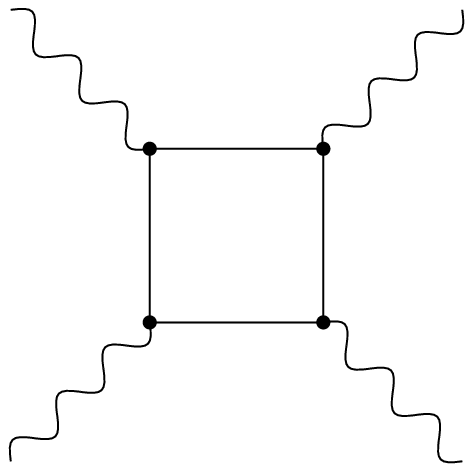,height=4cm,width=4cm}
\caption{Charged particle box diagram contributing to photon-photon scattering.}
\end{center}
\end{figure}

Direct evaluation of such a diagram is straightforward
but exceedingly tedious.  Nevertheless, the form of the result in the case
of scattering of photons with energy much smaller than the mass of the charged particles
is clear from the feature that it must be representable in terms of a simple local
interaction.  Since photon-photon scattering is involved, the effective Lagrangian
must be quartic in the vector potential and, since it must be gauge invariant, only 
the field tensors $F_{\mu\nu}=\partial_\mu A_\nu-\partial_\nu A_\mu$, its dual 
$\tilde{F}_{\mu\nu}={1\over 2}\epsilon_{\mu\nu\alpha\beta}F^{\alpha\beta}$ and 
their derivatives can be utilized.  Finally, the effective Lagrangian must be a Lorentz 
scalar and be parity even---{\it i.e.}, to lowest order it must have the form
\begin{equation}
{\cal L}_{eff}=\left({\alpha\over m^2}\right)^2\left[c_1(F_{\mu\nu}F^{\mu\nu})^2
+c_2(F_{\mu\nu}\tilde{F}^{\mu\nu})^2\right]
\end{equation}   
where we expect $c_1,c_2\sim {\cal O}(1).$  In fact explicit evaluation yields
\begin{eqnarray}
c_1&=&{7\over 1440},\quad c_2={1\over 1440}\qquad S=0\nonumber\\
c_1&=&{1\over 90},\quad c_2={7\over 360}\,\,\qquad S=1/2
\end{eqnarray}
which are the well-known Euler-Heisenberg results.\cite{euh}  Diagramatically this form 
corresponds to reduction of the box graph to a point interaction in accord with 
our discussion above.

\begin{figure}
\begin{center}
\epsfig{file=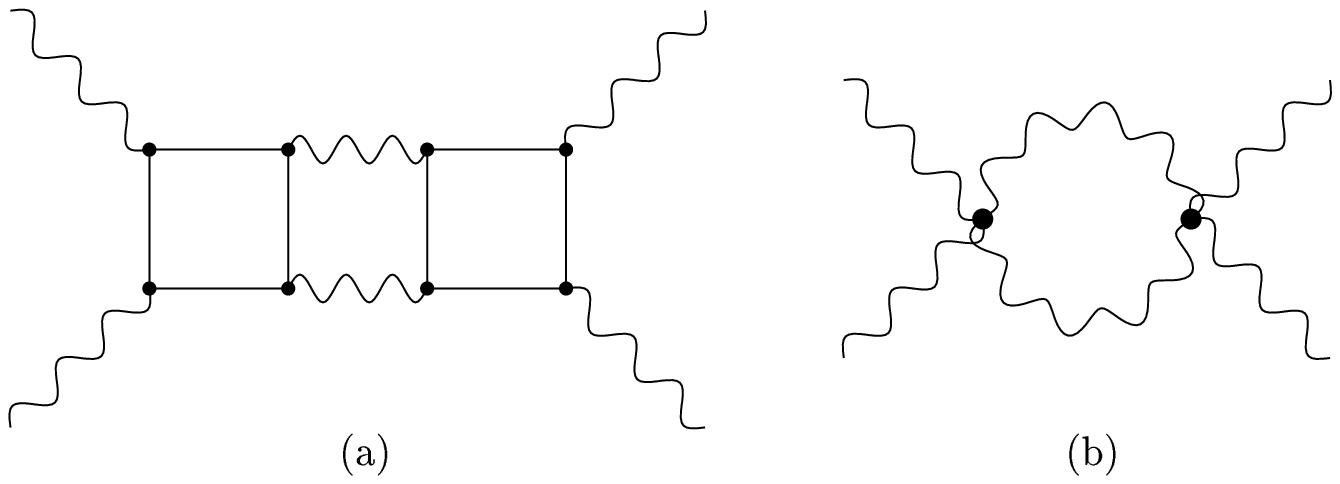,height=4cm,width=10cm}
\caption{A two-box QED diagram and its effective action analog.}
\end{center}
\end{figure}

An interesting and instructive 
situation arises associated with the diagram shown in figure 3a, wherein
{\it two} box diagrams are encoded.\cite{jfd}  In this case, by taking a cut across 
the two photons
in the middle of the graph unitarity requires a non-zero imaginary component---the 
diagram is complex!  However, such a complex number cannot be used as the coefficient
of a local interaction term, as it would violate hermiticity.  The resolution of 
this apparent paradox is that our Euler-Heisenberg contact interaction {\it itself} 
should be iterated in order to give the correct representation of the low energy field 
theory---{\it cf. } Figure 3b.  The (complex) low energy---long distance
component of this graph is guaranteed to reproduce the (complex) low energy piece of
the full graph Figure 3a.  However, an additional problem appears to surface when
the effective field theory loop in figure 3b is actually calculated, since it will
have the basic form
\begin{equation}
J(q)\sim\int{d^4s\over (2\pi)^4}{\alpha^2\over m^4}{q^2s\cdot(s+q)q^2s\cdot(s+q)\over s^2
(s+q)^2},
\end{equation}       
where $q$ represents an external momentum.  Hence $J(q)$ is divergent---in dimensional
regularization it will have the form
\begin{equation}
J(q)\sim{\alpha^4q^8\over m^8}\times({1\over d-4}-\gamma+\ldots)\label{eq:jjj}
\end{equation}
---while the
QED diagram which it is supposed to represent is finite.  

The resolution of this apparent paradox is that {\it only the low energy component} 
of the loop integration is
guaranteed to match in the full and effective loop diagrams.  This assures that the
long distance pieces, including the imaginary part discussed above, will agree.
As for the short distance piece, we must requere that the full effective theory
contain a {\it complete set} of possible local interactions----it must be of 
the form
\begin{equation}
{\cal L}_{eff}=-{1\over 4}F^2+{\alpha^2\over m^4}(F^2)^2+{d_1\over m^6}F^2\Box F^2
+{d_2\over m^8}F^2\Box^2 F^2+\ldots
\end{equation}      
Here the coefficient of the term in $(F^2)^2$ yields simply the Euler-Heisenberg
Lagrangian, while the remaining terms $d_1,d_2$,etc. are yet to be determined and 
are found by matching to the values calculated in the full theory.  The solution to
our dilemma is clear then---the divergence in Eq. \ref{eq:jjj} is simply combined
with the phenomenological coefficient $d_2$ before 
matching to the full theory. 

\subsection{Superconductivity}

Our last example before undertaking our goal of developing an effective
field theory for low energy QCD is that of superconductivity.  In this case
the underlying dynamics is associated with that of interactions of electrons
with the lattice and is described in terms of electron-phonon interactions.
Equivalently one can integrate out the lattice and write things in terms of
an effective interaction involving only electron degrees of freedom.  In this 
case it was shown in the 1950's by Cooper that there develops an attractive
interaction between electron pairs near the Fermi surface 
having opposite spins and momenta.  One can think of this qualitatively in 
terms of the presence of the first electron attracting the lattice atom, which
via screening then makes it easier for the second electron to be present.  In
any case one has then the effective interaction
\begin{equation}
{\cal L}_{eff}=\sum_{\vec{k},s}\psi_s^\dagger(\vec{k})
(i{\partial\over \partial t}
-{\vec{k}^2\over 2m}-\mu)\psi_s(\vec{k})+g\sum_{\vec{k},\vec{k}'}
\psi_\uparrow^\dagger(\vec{k})\psi_\downarrow^\dagger(-\vec{k})
\psi_\downarrow(-\vec{k}')\psi_\uparrow(\vec{k}')
\end{equation}
where $\mu$ is the chemical potential.  The usual (BCS) approach at this 
point is to solve this system using many body techniques, which leads to
a ground state involving a lowered energy due to condensation of these 
paired electron states and
which is separated by an energy gap $E_g$ between itself and its lowest 
excited state.  It is this "bosonization" property and energy gap which then
leads to superconductivity.

Unfortunately, this procedure is somewhat complicated and the physics is
not so easy to pick out from the detailed dynamical calculations which are
involved.  One can ameliorate this problem, however, by writing things in
terms of a two-electron state $\phi$ and integrating out the
electron states to yield an effective interaction.  Roughly we are doing
can be seen via
\begin{eqnarray}
W&=&\int[d\psi][d\psi^\dagger][d\phi]\delta[\phi-\psi\psi]\exp i\int d^3x
{\cal L}(\psi^\dagger,\phi)\nonumber\\
&=&\int [d\phi]\exp i\int d^3 {\cal L}_{eff}[\phi]
\end{eqnarray}
and diagrammatically the process can be represented as in Figure ???, where
one uses imaginary time techniques in order to produce results at fixed
temperature.  The result of this process is the effective Hamiltonian
\begin{equation}
{\cal H}_{eff}=-a(T)\phi^\dagger {D^2\over 2m^*}\phi
+b(T)\phi^\dagger\phi+c(T)(\phi^\dagger\phi)^2\label{eq:lg}
\end{equation}
where here $iD=-i\vec{\nabla}-e^*\vec{A}$ is the covariant derivative,
$m^*=2m,\,e^*=2e$ are the effective mass, charge of the electron pairs,
and $a(T),b(T),c(T)$ are calculable functions of the temperature.  However,
while $a(T),c(T)$ are positive and monotonic, $b(T)$ has the form $\ln
T/T_c$ where $T_c$ is a critical temperature defined in terms of the
coupling constant $g$.  Then above this temperature, the effective potential
is quartic in shape with a minimum at $\phi=0$.  However, for $T<T_c$ the
shape of the potential becomes as seen in Figure and a minimum develops at
a nonzero value of $\phi$, corresponding to the existence of a superconducting
state.  Also, we observe that Eq. \ref{eq:lg} has the form of the well
known Landau-Ginzburg Hamiltonian, in which the physics of superconductivity is
easily seen.

\subsection{QCD at Last}

After this introduction we finally turn to our goal of QCD.  I was a 
student in the 1960's and at that time our holy grail was to attempt to
find a renormalizable field theory which describes all particle interactions 
with the same sort of success as quantum electrodynamics (QED).  In 1967
we went part of the way with development of the Weinberg-Salam theory, which
incorporated the weak interaction as a sibling to the electromagnetic.
Because the interaction was weak it could be treated via the same perturbative
techniques as could its electromagnetic kin and what has resulted is an
extremely successful description of all weak and electromagnetic processes.

For the strong interactions a renormalizable picture has
also been developed---quantum chromodynamics or QCD. The theory is, of course,
deceptively simple on the surface.
Indeed the form of the Lagrangian\footnote{Here the covariant derivative is
\begin{equation}
i D_{\mu}=i\partial_{\mu}-gA_\mu^a {\lambda^a \over 2} \, ,
\end{equation}
where $\lambda^a$ (with $a=1,\ldots,8$) are the SU(3) Gell-Mann matrices,
operating in color space, and the color-field tensor is defined by
\begin{equation}
G_{\mu\nu}=\partial_\mu  A_\nu -  \partial_\nu  A_\mu -
g [A_\mu,A_\nu]  \, ,
\end{equation} }
\begin{equation}
{\cal L}_{\mbox{\tiny QCD}}=\bar{q}(i  {\not\!\! D} - m )q-
{1\over 2} {\rm tr} \; G_{\mu\nu}G^{\mu\nu} \, .
\end{equation}
is elegant, and the theory is renormalizable.  So why are we not
satisfied?  While at the very largest energies, asymptotic freedom 
allows the use of perturbative
techniques, for those who are interested in making contact with low energy 
experimental findings there exist at least three fundamental difficulties:
\begin{itemize}
\item [i)] QCD is written in terms of the "wrong" degrees of 
freedom---quarks and
gluons---while low energy experiments are performed with hadronic bound states;

\item [ii)] the theory is non-linear due to gluon self interaction;

\item[iii)] the theory is one of strong coupling---$g^2/4\pi\sim 1$---so that 
perturbative methods are not practical.
\end{itemize}
So what can be done.  Using our example of superconductivity we have seen
how the complicated BCS interaction in terms of spin-correlated electrons
can be replaced by a relatively simple effective (Landau-Ginzburg) interaction
in terms of a bound electron pair wavefunction $\phi$, wherein the physics is
clarified.  In the same way we shall try to do the same thing for QCD,
replacing the relatively complicated interaction in terms of quark and gluon
degrees of freedom by a simple form in terms of its $\bar{q}q$ and
$qqq$ bound states---{\it i.e.} mesons and baryons.  We shall accomplish
this by exploting the chiral symmetry of the QCD interaction. 

\section{Symmetry and Symmetry Breaking}
\subsection{Symmetry}
The best definition of symmetry for our purposes
is probably that due to the mathematician Herman Weyl who
said that a system is symmetric when one can do something to it and,
after making this change, the system looks the same as it did before.\cite{4}
The importance of symmetry in physics is due to an important
result---Noether's theorem---which connects
each symmetry of a system with a corresponding conserved current and
conservation law.\cite{5}  Familiar examples include:
\begin{itemize}
\item[--] ${\cal L} $ invariant under translation $\rightarrow$ momentum
conservation \par
\item[--] ${\cal L}$ invariant under time translation $\rightarrow$
energy conservation \par
\item[--] ${\cal L}$ invariant under rotation $\rightarrow$ angular
momentum conservation
\end{itemize}
and associated with each such invariance there is in general a related current 
$j_\mu$ which is conserved---{\it i.e.} $\partial^\mu j_\mu=0$.  This 
guarantees that the associated charge will be time-independent, since
\begin{equation}
{dQ\over dt}=\int d^3x{\partial j_0\over \partial t}=-\int d^3x \vec{\nabla}\cdot\vec{j}
=\int_{\rm surf}d\vec{S}\cdot\vec{j}=0 
\end{equation}
where we have used Gauss' theorem and the assumption that any fields are local.
Given a specific field theory, we can identify the Noether
currents via standard techniques.  Suppose that the
Lagrangian is invariant under the transformation
$\phi\longrightarrow \phi+\varepsilon f(\phi)$---{\it i.e.}
\begin{eqnarray}
0 & =& {\cal L} (\phi + \varepsilon f, \partial_{\mu} \phi +\varepsilon
\partial_{\mu} f) -{\cal L}(\phi, \partial_{\mu}\phi) \nonumber \\
&=&  \varepsilon f {\delta {\cal L} \over \delta \phi}
+\varepsilon \partial_{\mu} f {\delta {\cal L}\over \delta (
\partial_{\mu}\phi)}
=\varepsilon \partial_{\mu} \left( f {\delta {\cal L} \over
\delta(\partial_{\mu}\phi)}\right)  \, .
\end{eqnarray}
so that we can identify the associated conserved current as \footnote{Note:
This is often written in an alternative fashion by introducing a
{\it local} transformation
 $\varepsilon = \varepsilon(x)$, so that the Lagrangian transforms as
\begin{equation}
 {\cal L}(\phi,\partial_{\mu}\phi) \rightarrow
{\cal L} (\phi+\varepsilon f, \partial_{\mu} \phi +
\varepsilon \partial_{\mu} f +f \partial_{\mu}
\varepsilon) \, .
\end{equation}
Then
\begin{equation}
 {\delta {\cal L} \over \delta(\partial_{\mu}
\varepsilon)} = f {\delta {\cal L} \over \delta
(\partial_{\mu}\phi)} \,  ,
\end{equation}
so that the Noether current can also be written as
\begin{equation}
 j^{\mu} = {\delta {\cal L} \over \delta(
\partial_{\mu}\varepsilon)} \,  .
\end{equation}}
\begin{equation}
 j^{\mu} = f {\delta{\cal L} \over
 \delta(\partial_{\mu}\phi)} \, ,
\end{equation}
Since in quantum mechanics the time development of an operator 
$\hat{Q}$ is given by
\begin{equation}
{d\hat{Q}\over dt}=i[\hat{H},\hat{Q}]
\end{equation}
we see that such a conserved 
charge must commute with the Hamiltonian.  Now in general the
vacuum (or lowest energy state) of such a theory, which satisfies 
$\hat{H}|0>=E_0|0>$,
is unique and has the property $\hat{Q}|0>=|0>$ 
since $\hat{H}(\hat{Q}|0>)=\hat{Q}(\hat{H}|0>)=E_0\hat{Q}|0>$.  
In this case we say that the symmetry is realized in a Wigner-Weyl fashion and
there will exist in general a set of degenerate excited states which mix with
each other under application of the symmetry charge.  A familiar example of a
Wigner-Weyl symmetry is isospin or SU(2) invariance.  Because this is an 
(approximate) symmetry of the Hamiltonian, 
particles appear in multiplets such as $p,n$ or $\pi^+\pi^0\pi^-$ having
the same spin-parity and (almost) the same mass and transform into one 
another under
application of the isospin charges $\vec{I}$.  However, this is not the only 
situation which occurs in nature.  It is also possible (and in fact 
often the case) 
that the ground state of a system does {\it not} have the same symmetry 
as does the 
Hamiltonian, in which case we say that the symmetry is realized in
a Nambu-Goldstone fashion and is "spontaneously broken".  This
phenomenon is actually a familiar one from classical mechanics and was
first studied by Euler in the context of a rod under compression and by
Jacobi in the context of a rotating earth.

In fact the idea exact symmetry and therefore exact conservation laws is
not the usual one in nature.  Rather it is common to have to deal with
approximate symmetries which would be exact only in some
hypothetical universe which is not our own---in our world
such symmetries will
be seen to be broken in some fashion.  In spite of this,
such broken symmetries are
of great
importance and by their study we will be able to learn much
about the underlying interactions.
\subsection{Symmetry Breaking} \par
In general there exist in physics only three possible mechanisms for symmetry
breaking
\begin{itemize}
\item[--] explicit symmetry breaking
\item[--] spontaneous symmetry breaking
\item[--] quantum mechanical symmetry breaking
\end{itemize}
and in this section we study examples of each:\\

{\bf Explicit Symmetry Breaking}\\

First consider a simple harmonic oscillator of frequency $\omega_0$
described by the Lagrangian

\begin{equation}
L= {1\over 2}m {\dot x}^2-{1\over 2} m \omega_0^2 x^2 \, .
\end{equation}
which is explicitly invariant under spatial
inversion---$x\rightarrow -x$---since
\begin{equation}
V_0(x)= -{1 \over 2} m\omega_0^2 x^2=-{1 \over 2} m\omega^2
(-x)^2=V_0(-x)
\end{equation}
Thus it is clear from symmetry considerations that the equilibrium
location $x_E$, which is determined
by the condition $[\partial L/ \partial x](x_E)=0$,
must occur at $x_E=0$, since the equilibrium position
should also manifest this symmetry.

Now, however, consider what happens if we add an term $V_1(x)=\lambda x$
{\it i.e.} a constant force, to the Lagrangian.   The new Lagrangian
is
\begin{equation}
L={1\over 2}m\dot{x}^2-{1\over 2}m\omega_0^2x^2+\lambda x
\end{equation}
which describes a displaced oscillator.  This new Lagrangian is {\it not}
invariant under spatial inversion, and consequently the new equilibrium
location---$x_E=\lambda / m\omega^2\neq 0$---is no longer required
to be at the origin.  This is an example of {\it explicit symmetry breaking}
wherein the symmetry violation is manifested in the Lagrangian itself.\\

{\bf Spontaneous Symmetry Breaking}\\

As our second example, consider a hoop rotating in the earth's gravitational
field about a vertical axis.\cite{6}  Attached to the hoop is a bead which can
slide
along the circumference without friction.  The lagrangian $L$ for the system
is then
\begin{equation}
L= {1\over 2} m (R^2 {\dot  \theta}^2+\omega^2 R^2 \sin^2\theta)
+m g R \cos \theta \, ,
\end{equation}
where $\theta$ measures the angular displacement of the bead from the nadir.
$L $ is clearly symmetric under the angular parity transformation
$L(\theta)=L(-\theta)$, but the equilibrium condition for the bead
is found to be
\begin{equation}
{\partial L\over \partial \theta }=m \omega^2 R^2\sin\theta(\cos\theta-
{ g\over {\omega^2 R}})=0  \, .
\end{equation}
which is somewhat more complex than the displaced oscillator considered above.
For slow rotation---{\it i.e } for $\omega^2<{g\over R}$, we have
$\cos\theta-{g\over{\omega^2 R}} \neq 0$,
so that the ground (equilibrium) state configuration is given by $\theta_E=0$
as expected from symmetry considerations.  However, if we proceed to higher
angular velocities such that $\omega^2 > {g\over R}$ then
the bead finds equilibrium at $\theta_E=\pm\cos^{-1}{g\over \omega^2 R}$,
where the choice of + vs. - is {\it not} determined by the physics but
rather by the
history of motion of the system as the critical angular velocity was
reached.  Note that neither of these equilibrium positions exhibits
the symmetry of the underlying potential, which is invariant
under the exchange of $\theta$ and $-\theta$.  This is an example of
{\it spontaneous symmetry breaking}, wherein the
Lagrangian of a system possess a symmetry, but this symmetry is broken
by the ground (equilibrium) state of the system.\\

{\bf Quantum Mechanical Symmetry Breaking}\\

The third type of symmetry breaking is the least familiar to most
physicists because it
has no classical analog.  It is called ``quantum mechanical"
or ``anomalous" symmetry breaking and occurs when the classical
Lagrangian of a system possesses a symmetry, but the symmetry broken
in the process of quantization.
As the simplest example and the only one (of which I am aware) that does not
involve quantum field theory---just quantum mechanics!---consider a
free particle, for which the stationary state Schr\"odinger equation is\cite{7}

\begin{equation}
-{1\over 2 m} \nabla^2 \psi= E \psi\equiv {k^2\over 2m}\psi  \, ,
\end{equation}
A partial wave solution in polar coordinates is
\begin{equation} \psi(\vec{r})={1\over r} \chi_k(r) P_l (\cos\theta) \, ,
\end{equation}
where $\chi_k(\vec{r})$ satisfies the radial Schr\"odinger equation

\begin{equation} \left(-{d^2\over d r^2}+ {l(l+1)\over r^2} + k^2\right)
\chi_k(r)=0\, .\end{equation}
Here the central piece in the above differential operator is the
well-known ``centrifugal potential."  By inspection the radial Schr\"odinger
equation is invariant under a ``scale transformation"

\begin{equation} r \rightarrow \lambda r \quad\quad k \rightarrow {1\over
\lambda }k\, .\end{equation}
This {\it scale invariance} has an important physical consequence, which can
be seen if we expand a plane wave solution in terms of incoming and outgoing
partial waves

\begin{equation} e^{ikz} \stackrel{r \rightarrow \infty}{\longrightarrow}
{1\over 2 ikr}
\sum_l (2l+1) P_l (\cos\theta)\left(e^{ikr} -e^{-i(kr-l\pi)} \right)\,
,\end{equation}
We observe that in each partial wave the incoming and outgoing component
of the wavefunction differ by the centrifugal phase shift $l\pi$.
This phase shift must be {\it independent} of energy via scale invariance.

If we place the free particle in a potential $V(\vec {r})$ then the
scale invariance is broken.  The corresponding wave function expanded in
partial waves then becomes
\begin{equation}
\psi^{(+)}(\vec r) \stackrel{ r \rightarrow \infty} {\longrightarrow}
{1\over 2 i k r} \sum_l (2l+1) P_l (\cos \theta)
(e^{i(kr+2\delta_l (k))}-e^{-i(kr-l\pi)} )
\end{equation}
Usually this is written as
\begin{equation}
\psi^{(+)}(\vec r)=e^{ikx}+{e^{ikr}\over r}f_k(\theta)
\end{equation}
where the scattering amplitude is defined by

\begin{equation}  f_k(\theta)= \sum_l(2l+1) {e^{2i\delta_l(k)}-1 \over 2ik}
P_l(\cos \theta)\, .\end{equation}
Of course, the phase shifts $\delta_l(k)$ of various angular momenta $l$ now
depend on energy, but this is to be expected since the scale invariance no
longer obtains.

One can generalize the scattering formalism to two dimensions, in which case
we obtain for the scattering wave function

\begin{equation}\psi^{(+)}(\vec r)  \stackrel{ r \rightarrow \infty}
{\longrightarrow}
e^{ikz} +{1\over \sqrt{r}} e^{i(kr+{\pi \over 4})} f_k(\theta)\end{equation}
and for the scattering amplitude

\begin{equation} f_k(\theta)=-i\sum^{\infty}_{m=-\infty} {e^{2i\delta_m(k)}-1
\over
\sqrt{2 \pi k} }e^{im\theta} \, \end{equation}
where we expand in terms of exponentials $e^{im\theta}$ rather than Legendre
polynomials.  What is {\it special} about two dimensions is that it
is possible to introduce
a {\it scale invariant} potential
\begin{equation} V(\vec r)=g\delta^2(\vec r)\end{equation}
The associated differential scattering cross section is found to be\cite{8}
\begin{equation}{d \sigma \over d \Omega} \propto {\pi \over 2 k} {1\over
(\ln {k\over \mu}^2)} \, .\end{equation}
which is somewhat of a surprise.  Indeed since the cross section is isotropic,
the scattering is pure $m=0$, corresponding to a phase shift

\begin{equation}  \cot \delta_0(k)={1\over \pi} \ln {k^2\over \mu^2}-{2\over
g}\, ,\end{equation}
which depends on $k$---scale invariance has been broken
as a result of quantization.  Although this should not be completely unexpected
(indeed while at the classical level non-zero impact parameter means
no scattering, in quantum mechanics this is not the case because of the
non-zero deBroglie wavelength), still the ``physics" of this result is not
completely clear.

\medskip

{\bf Chiral Symmetry}\par\bigskip

In order to understand the relevance of spontaneous symmetry breaking 
within  QCD,
we must introduce the idea of "chirality," defined by the operators
\begin{equation} \Gamma_{L,R} = {1\over 2}(1\pm\gamma_5)={1\over 2}
\left( \begin{array}{c c }
1 & \mp 1 \\
\mp 1 & 1
\end{array}\right)
\end{equation}
which project left- and right-handed components of the Dirac wavefunction
via
\begin{equation} \psi_L = \Gamma_L \psi \qquad \psi_R=\Gamma_R
\psi \quad\mbox{with}\quad \psi=\psi_L+\psi_R \end{equation}
In terms of these chirality states the quark component of the QCD Lagrangian
can be written as
\begin{equation} \bar{q}(i\not\! \! D-m)q=\bar{q}_Li\not \! \! D q_L +
\bar{q}_Ri\not\!\! D q_R -\bar{q}_L m q_R-\bar{q}_R m
q_L \end{equation}
The reason that these chirality states are called left- and right-handed can
be seen by examining helicity eigenstates of the free Dirac equation.  In the
high energy (or massless) limit we note that 
\begin{equation}
 u(p)= \sqrt{{E+m\over 2E}}
\left( \begin{array}{c }
\chi \\  {\vec{\sigma}\cdot\vec{p} \over E+m}\chi
\end{array}\right)
\stackrel{E \gg m}{\sim} \sqrt{{1\over 2}}
\left( \begin{array}{c }
\chi \\  \vec{\sigma}\cdot\hat{p} \chi
\end{array}\right)
\end{equation}
Left- and right-handed helicity eigenstates then can be identified as
\begin{equation}
u_L(p)  \sim  \sqrt{1\over 2}
\left( \begin{array}{c}
\chi \\ -\chi
\end{array} \right),\qquad
u_R(p)  \sim  \sqrt{1\over 2}
\left( \begin{array}{c}
\chi \\ \chi
\end{array} \right)
\end{equation}
But 
\begin{eqnarray}
 \Gamma_L u_L= u_L && \Gamma_R u_L=0 \nonumber \\
 \Gamma_R u_R= u_R && \Gamma_L u_R =0
\end{eqnarray}
so that in this limit chirality is identical with helicity---
\[ \Gamma_{L,R} \sim \mbox{helicity!} \]

With this background, we now return to QCD and observe that in the limit as  
$m\rightarrow0$ 
\begin{equation} {\cal L}_{\rm QCD}\rightarrow\bar{q}_L i \not\!\! D q_L +
\bar{q}_R i \not\!\! D q_R \end{equation}
would be invariant under {\it independent} global
left- and right-handed rotations
\begin{equation}
q_L  \rightarrow \exp (i \sum_j \lambda_j\alpha_j)
q_L,\qquad
q_R  \rightarrow \exp (i\sum_j \lambda_j \beta_j)
q_R
\end{equation}
(Of course, in this limit the heavy quark component is also invariant, but
since $m_{c,b,t} >> \Lambda_{\rm QCD}$ it would be silly to consider this as
even an approximate symmetry in the real world.)  This invariance is called
$SU(3)_L \bigotimes SU(3)_R$ or chiral $SU(3)\times SU(3)$.  Continuing
to neglect the light quark masses,
we see that in a chiral symmetric world one would expect to have have sixteen---eight
left-handed and eight right-handed---conserved Noether currents
\begin{equation} \bar{q}_L\gamma_{\mu} {1\over 2} \lambda_i q_L \, ,
\qquad \bar{q}_R\gamma_{\mu}{1\over 2}\lambda_i
q_R \end{equation}
Equivalently, by taking the sum and difference we would have eight conserved vector and
eight conserved axial vector currents
\begin{equation}
V^i_{\mu}=\bar{q}\gamma_{\mu} {1\over 2}
\lambda_i q,\qquad
A^i_{\mu}=\bar{q}\gamma_{\mu}\gamma_5
 {1\over 2} \lambda_i q
\end{equation}
In the vector case, this is just a simple generalization of isospin (SU(2)) invariance 
to the case of SU(3).  There exist 
{\it eight} ($3^2-1$) time-independent charges
\begin{equation} F_i=\int d^3 x V^i_0(\vec{x},t) \end{equation}
and there exist various supermultiplets of particles having identical spin-parity and
(approximately) the same mass in the configurations---singlet, octet, decuplet, 
{\it etc.} demanded by SU(3)invariance.

If chiral symmetry were realized in the conventional fashion one would
expect there also to exist corresponding nearly degenerate
but {\it opposite} parity states generated
by the action of the time-independent axial charges
$F^{5}_i= \int d^3 xA^i_0(\vec{x},t)$
on these states.  Indeed since
\begin{eqnarray}
H|P\rangle &= & E_P|P\rangle \nonumber \\
H(Q_5|P\rangle)&=&Q_5(H|P\rangle)
=  E_P(Q_5|P\rangle)
\end{eqnarray}
we see that $Q_5|P\rangle$ must also be an eigenstate of the Hamiltonian
with the same eigenvalue as $|P>$, which would seem to require the existence of
parity doublets.  However, experimentally this does not appear to be the
case.  Indeed although the $J^p={1\over 2}^+$ nucleon has a mass of about 1 GeV, the
nearest ${1\over 2}^-$ resonce lies nearly 600 MeV higher in energy.  Likewise in
the case of the $0^-$ pion which has a mass of about 140 MeV, the nearest 
corresponding $0^+$ state (if it exists at all) is nearly 700 MeV or so higher 
in energy. 
\bigskip
\subsection{Goldstone's Theorem}

One can resolve this apparent paradox by postulating that parity-doubling is
avoided because the axial symmetry is
{\it spontaneously broken.}  Then according to a theorem due to Goldstone,
when a continuous symmetry is broken in this fashion there must also be
generated a {\it massless} boson having the quantum
numbers of the broken generator---in this case a pseudoscalar---and when the
axial charge acts on a single particle eigenstate one does not get a single
particle eigenstate of opposite parity in return.\cite{gold} Rather one generates
one or more of these massless pseudoscalar bosons
\begin{equation} 
Q_5|P\rangle \sim |Pa \rangle +\cdots 
\end{equation}
and the interactions of such "Goldstone bosons" to each other and to other particles is
found to vanish as the four-momentum goes to zero.  

In order to see how the corresponding situation develops in QCD, it is useful to
study a simple pedagogical example---a scalar field theory\cite{burg} 
\begin{equation}
{\cal L}=\partial_\mu\phi^*\partial^\mu\phi-V(\phi^*\phi)\quad{\rm with}\quad
V(x)={\lambda\over 4}(x-{\mu^2\over\lambda})^2
\end{equation} 
which is obviously invariant under the global 
U(1) (phase) transformation $\phi\rightarrow
e^{i\alpha}\phi$.  The vacuum (lowest energy) state of the system can be found by
minimizing the Hamiltonian density
\begin{equation}
{\cal H}=\dot{\phi}^*\dot{\phi}+\vec{\nabla}\phi^*\cdot\vec{\nabla}\phi+V(\phi^*\phi)
\end{equation} 
Since this is the sum of positive definite terms, the vacuum state is easily seen to
be $\phi=v=\mu/\sqrt{\lambda}$, where U(1) symmetry has been used in order to
choose $v$ as real.  Of course, once this is done the U(1) symmetry is 
broken---spontaneous symmetry breaking has takne place---and Goldstone's 
theorem is applicable.

In order to see how this comes about we select as independent fields the real and
imaginary components of $\phi$---$\rho\equiv\sqrt{2}({\rm Re}\phi-v),\,\,\chi\equiv
\sqrt{2}{\rm Im}\phi$---in terms of which the Lagrangian density becomes
\begin{equation}
{\cal L}={1\over 2}\partial_\mu\rho\partial^\mu\rho+
{1\over 2}\partial_\mu\chi\partial^\mu\chi-{1\over 2}\mu^2\rho^2
-{\lambda\over 16}(\rho^2+\chi^2)^2-{\mu\sqrt{\lambda}\over 2\sqrt{2}}\rho(\rho^2+\chi^2)
\end{equation} 
We observe that the field $\chi$ is massless---this is the Goldstone mode---while the
field $\rho$ has a mass $\mu$.  The Noether current
\begin{equation}
j_\mu=-i(\phi^*\partial_\mu\phi-\partial_\mu\phi^*\phi)=\sqrt{2}v\partial_\mu\chi
+\rho\partial_\mu\chi-\chi\partial_\mu\rho
\end{equation}
possesses a nonzero matrix element between $\chi$ and the vacuum
\begin{equation}
<\chi(p)|j_\mu|0>=\sqrt{2}vip_\mu e^{-ip\cdot x}\label{eq:qq}
\end{equation}  
provided that $v\neq 0$. Also there exist complicated self interactions as
well as mutual interactions between $\rho$ and $\sigma$.  However, if we calculate
the tree-level amplitude for $\rho\chi$ scattering, using the diagrams illustrated 
in figure 4 we find
\begin{eqnarray}
{\rm Amp}(\rho(q)+\chi(p)&\rightarrow&\rho(q')+\chi(p'))=
{\lambda\over 2}+{3\over 2}\lambda^2v^2{1\over (p+p')^2-\mu^2}\nonumber\\
&+&{1\over 2}\lambda^2v^2\left({1\over (p+q)^2}+{1\over (p-q')^2}\right),
\end{eqnarray}
and in the soft momentum limit for the Goldstone bosons---$p,p'\rightarrow 0$---we
find that 
\begin{equation}
\lim_{p,p'\rightarrow 0}{\rm Amp}={\lambda\over 2}-{3\over 2}{\lambda^2v^2\over \mu^2}
+{\lambda^2v^2\over \mu^2}=0\label{eq:kk}
\end{equation}
{\it i.e.}, the amplitude vanishes, as asserted above.  

\begin{figure}
\begin{center}
\epsfig{file=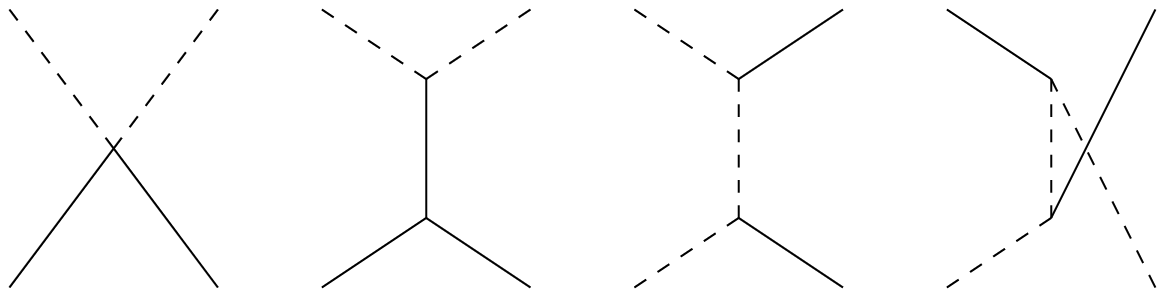}
\caption{Toy model $\rho\chi$ scattering diagrams.}
\end{center}
\end{figure}

Thus our toy model certainly has all the right stuff, but our representation of the
fields is not the optimal one in order to display the Goldstone properties.  Instead
it is advantageous to utilize a polar co-ordinate representation in which the Goldstone
mode appears in the guise of a local U(1) 
transformation---$\phi=(v+\sqrt{1\over 2}\xi)\exp i\theta/ v\sqrt{2}$---whereby 
the Lagrangian density assumes the form
\begin{equation}
{\cal L}={1\over 2}\partial_\mu\xi\partial^\mu\xi+{1\over 2}
(1+{1\over \sqrt{2}}{\xi\over v})^2\partial_\mu\theta\partial^\mu\theta
-{1\over 2}\mu^2\xi^2-{\mu\lambda\over 2\sqrt{2}}
\xi^3-{\lambda\over 16}\xi^4
\end{equation}
We see in this form that $\theta$ is the massless Goldstone field, while the field 
$\xi$ has mass $\mu$.  The Noether current 
\begin{equation}
j_\mu=\sqrt{2}v
(1+\sqrt{1\over 2}{\xi\over v})^2\partial_\mu\theta
\end{equation}
clearly has a nonzero vacuum-Goldstone matrix element which agrees with Eq. \ref{eq:qq}.
However, what is particularly useful about this representation is the feature that
the Goldstone modes couple only through derivative coupling.   Thus the feature that 
any such interactions must vanish in the soft momentum limit is displayed explicitly,
making Eq. \ref{eq:kk} trivial.   
   
Now back to QCD:  According to Goldstone's argument, one would expect 
there to exist eight {\it massless} pseudoscalar states---one for each 
spontaneously broken SU(3) axial generator, which would be the Goldstone bosons of QCD.
Examination of the particle data tables reveals, however, that no such massless $0^-$ particles
exist.  There do exist eight $0^-$ particles---$\pi^\pm,\pi^0,K^\pm,K^0,\bar{K}^0\eta$
which are much lighter than their hadronic siblings.  However, these states are
certainly not massless and this causes us to ask what has gone wrong with what 
appears to be rigorous reasoning.  The answer
is found in the feature that our discussion thus far has neglected the piece
of the QCD Lagrangian which is associated with quark mass and can be
written in the form
\begin{equation} {\cal L}_{\mbox{QCD}}^m=-(\bar{u}_L u_R+\bar{u}_R
u_L) m_u-(\bar{d}_L d_R+\bar{d}_R d_L)m_d \end{equation}
Since clearly this term breaks the chiral symmetry---
\begin{eqnarray}
\bar{q}_L q_R &\rightarrow & \bar{q}_L \exp(-i
\sum_j \lambda_j\alpha_j) \times \exp(i\sum_j \lambda_j
\beta_j) q_R \nonumber \\
& \neq & \bar{q}_L q_R
\end{eqnarray}
---we have a violation of the conditions under which Goldstone's theorem aplies.  
The associated pseudoscalar bosons are {\it not} required to be massless
\begin{equation} m^2_G \neq 0 \end{equation}
but since their mass arises only from the breaking of the symmetry the
various "would-be" Goldstone masses are expected to be proportional to
the symmetry breaking parameters
\[ m^2_G \propto m_u,m_d,m_s \]
To the extent that such quark masses are small 
the eight pseudoscalar masses are not required to be massless, merely 
much lighter than other hadronic masses in the spectrum, as found in nature.

\subsection{Effective Chiral Lagrangian}

The existence of a set of particles---the pseudoscalar mesons---which are 
notably less massive than other hadrons suggests the possibility of generating
an effective field theory which correctly incorporates the chiral symmetry of the
underlying QCD Lagrangian in describing the low energy interactions of these
would-be Goldstone particles.  As found in our pedagogical example, and in a 
homework problem, this can
be formulated in a variety of ways, but the most transparent is done by including the
Goldstone modes in terms of the argument of an exponential 
$U=\exp(i\vec{\tau}\cdot\phi/v)$ such that under
the chiral transformations
\begin{eqnarray}
\psi_L & \rightarrow & L \psi_L \nonumber \\
\psi_R & \rightarrow & R \psi_R
\end{eqnarray}
we have
\begin{equation} U \rightarrow L U R ^{\dagger} \end{equation}
Then a form such as
\begin{equation} \mbox{Tr} \partial^{\mu} U \partial_{\mu} U^{\dagger}
\rightarrow \mbox{Tr} L \partial^{\mu} U R^{\dagger} R
\partial_{\mu} U^{\dagger} L^{\dagger} = \mbox{Tr} \partial^{\mu}
U \partial_{\mu} U^{\dagger}\,   
\end{equation}
is invariant under chiral rotations and can be used as part of the
effective Lagrangian.  However, this form is
also not one which we can use in order to realistically describe
Goldstone interactions in Nature since according to Goldstone's theorem
a completely invariant Lagrangian must also have zero pion mass, in
contradiction to experiment.  We {\it must} include a term which uses the
quark masses to generate chiral symmetry breaking and thereby non-zero pion
mass.  

We infer then that the {\it lowest order} effective chiral Lagrangian 
can be written as
\begin{equation}
 {\cal L}_2={v^2 \over 4} \mbox{Tr} (\partial_{\mu}U \partial^{\mu}
 U^{\dagger})+{m^2_{\pi}\over 4} v^2 \mbox{Tr} (U+U^{\dagger})\,  .\label{eq:abc}
\end{equation}
where the subscript 2 indicates that we are working at two-derivative order
or one power of chiral symmetry breaking---{\it i.e.} $m_\pi^2$.
This Lagrangian is also {\it unique}---if we expand to lowest order in
$\vec\phi$
\begin{equation}
\mbox{Tr}\partial_{\mu} U \partial^{\mu} U^{\dagger} =
\mbox{Tr} {i\over v} \vec{\tau}\cdot\partial_{\mu}\vec{\phi} \times
{-i\over v}\vec{\tau}\cdot\partial^{\mu}\vec{\phi}= {2\over v^2}
\partial_{\mu}\vec{\phi}\cdot \partial^{\mu}\vec{\phi}\,  ,
\end{equation}
we reproduce the free pion Lagrangian, as required,
\begin{equation}
 {\cal L}_2 ={1\over 2} \partial_{\mu}
\vec{\phi}\cdot \partial^{\mu} \vec{\phi} -{1\over 2} m^2_{\pi}
\vec{\phi}\cdot \vec{\phi} +{\cal O} (\phi^4) \,  .
\end{equation}

At the SU(3) level, including a generalized chiral symmetry breaking term,
there is even predictive power---one has
\begin{equation}
 {v^2\over 4} \mbox{Tr} \partial_{\mu} U \partial^{\mu} U^{\dagger}
=   {1\over 2} \sum_{j=1}^8 \partial_{\mu}
\phi_j\partial^{\mu}\phi_j +\cdots \nonumber\\
\end{equation}
\begin{eqnarray}
{v^2 \over 4} \mbox{Tr} 2 B_0 m ( U+ U^{\dagger})
& =&  \mbox{const.}
-{1\over 2} (m_u+ m_d)B_0 \sum_{j=1}^3 \phi^2_j \nonumber\\
&-&{1\over 4} (m_u+m_d+2m_s)B_0\sum_{j=4}^7 \phi^2_j
 -{1\over 6} (m_u+m_d +4m_s)B_0\phi^2_8  +\cdots \, \nonumber\\
&&
\end{eqnarray}
where $B_0$ is a constant and $m$ is the quark mass matrix. We can
then identify the meson masses as
\begin{eqnarray}
 m^2_{\pi} & =&  2\hat{m} B_0
\nonumber \\
 m_K^2 &=& (\hat{m} +m_s) B_0 \nonumber \\
m_{\eta}^2 & =& {2\over 3} (\hat{m} + 2m_s) B_0  \, ,
\end{eqnarray}
where $\hat{m}={1\over 2}(m_u+m_d)$ is the mean light quark mass.
This system of three equations is {\it overdetermined}, and we find by simple
algebra
\begin{equation}
3m_{\eta}^2 +m_{\pi}^2 - 4m_K^2 =0 \, \, .
\end{equation}
which is the Gell-Mann-Okubo mass relation and is well-satisfied
experimentally.\cite{gmo}

\bigskip

{\bf Currents }\par\bigskip

Since under a
\begin{equation}
\mbox{Vector, Axial transformation:}\,\,\,  \alpha_L  =\pm\alpha_R
\end{equation}
we have
\begin{equation}
U \rightarrow  \mbox{LUR}^{\dagger}  \stackrel{V}
{\simeq}  U + i\left[\sum_j \alpha_j\lambda_j, U
\right]
\stackrel{A}
{\simeq} U + i\left\{\sum_j \alpha_j\lambda_j, U
\right\}  \,  .
\end{equation}
which leads to the vector and axial-vector currents
\begin{equation}
\{V,A\}^k_{\mu}=  -i{v^2\over 4}\mbox{Tr}
\lambda^k(U^{\dagger}\partial_{\mu}U\pm U
\partial_{\mu}U^{\dagger})
\end{equation}

At this point the constant $v$ can be identified by use of the axial current.
In SU(2) we find
\begin{eqnarray}
U^{\dagger}\partial_{\mu}U-U\partial_{\mu}
U^{\dagger} &= & 2i{1\over v} \vec{\tau}
\cdot\partial_{\mu}\vec{\phi}+ \cdots
\end{eqnarray}
so that
\begin{equation}
A^k_{\mu}=  i{v^2\over 4} \mbox{Tr}\tau^k
2i{1\over v} \vec{\tau}\cdot\partial_{\mu}\vec{\phi}
+\cdots = -v \partial_{\mu}\phi^k+\cdots \, .
\end{equation}
If we set $k=1-i2$ then this represents the axial-vector component of the
$\Delta S=0$ charged weak current and
\begin{equation}
 A_{\mu}^{1-i2}= -v\partial_{\mu}\phi^{1-i2}
=-\sqrt{2}v \partial_{\mu}\phi^- \,  .
\end{equation}
Comparing with the conventional definition
\begin{equation}
\langle 0|A_{\mu}^{1-i2}(0)|\pi^+
(p)\rangle = i\sqrt{2}F_{\pi}p_{\mu} \, ,
\end{equation}
we find that, to lowest order in chiral symmetry, $v=F_\pi$, where
$F_\pi=92.4$ MeV is the pion decay constant.\cite{bhm}

Likewise in SU(2), we note that
\begin{equation} U^{\dagger} \partial_{\mu} U + U\partial_{\mu}
U^{\dagger}= {2i\over v^2}\vec{\tau}\cdot\vec{ \phi}
\times \partial_{\mu}\vec{\phi}+\cdots \, , \end{equation}
so that the {\it vector current} is
\begin{eqnarray}
V_{\mu}^k & = & -i {v^2\over 4}\mbox{Tr}\tau^k
{2i\over v^2}\vec{\tau}\vec{\phi}\times
\partial_{\mu}\vec{\phi}+\cdots \nonumber \\
& =& (\vec{\phi}\times\partial_{\mu}\vec{\phi})^k+\cdot \, .
\end{eqnarray}
We can identify $V_\mu^k$ as the (isovector) electromagnetic current by setting $k=3$
so that
\begin{equation}
 V_{\mu}^{\rm em}=\phi^+\partial_{\mu}\phi^-
-\phi^-\partial_{\mu}\phi^+ +\cdots
\end{equation}
Comparing with the conventional definition
\begin{equation}
\langle\pi^+(p_2)|V^{\rm em}_{\mu}(0)|
\pi^+(p_1)\rangle =F_1(q^2)(p_1+p_2)_{\mu}\,  ,
\end{equation}
we identify the pion form factor---$F_1(q^2)=1$. Thus to lowest order in
chiral symmetry the pion has unit charge but is pointlike and structureless.
We shall see below how to insert structure.

{\bf $\pi\pi$ Scattering}\par\bigskip

At two derivative level we can generate additional predictions by
extending our analysis to the case of $\pi\pi$ scattering.  Expanding
${\cal L}_2$ to order $\vec{\phi}^4$ we find
\begin{equation}
{\cal L}_2 : \phi^4={1\over 6v^2} \vec{\phi}^2\vec{\phi}\cdot \Box\vec{\phi} +
{1\over 2v^2} (\vec{\phi}\cdot\partial_{\mu}\vec{\phi})^2
 + { m^2_\pi \over 24 v^2} \vec{\phi}^4
\end{equation}
which yields for the pi-pi $T$ matrix
\begin{eqnarray}
T(q_a, q_b;q_c, q_d) & =& {1\over F_{\pi}^2} \left[ \delta^{ab}
\delta^{cd}(s-m^2_{\pi}) +\delta^{ab}\delta^{bd}(t-m_{\pi}^2)
+\delta^{ad}\delta^{bc} (u -m_{\pi}^2)\right]  \nonumber \\
&& -{1\over 3F^2_\pi}(\delta^{ab}\delta^{cd}+\delta^{ac}\delta^{bd}+\delta^{ad}
\delta^{bc})(q^2_a+q_b^2+q_c^2+q_d^2 - 4 m_{\pi}^2)
 \, . \nonumber\\
&&
\end{eqnarray}
Defining more generally
\begin{equation}
 T_{\alpha\beta;  \, \gamma\delta}(s,t,u)=A(s,t,u)
\delta_{\alpha\beta}
\delta_{\gamma\delta}+A(t,s,u)\delta_{\alpha\gamma}
\delta_{\beta\delta}+A(u,t,s)\delta_{\alpha\delta}\delta_{\beta\gamma} \, ,
\end{equation}
we can write the chiral prediction in terms of the more conventional isospin
language by taking appropriate linear combinations\cite{dsm}
\begin{eqnarray}
T^0(s,t,u)&=& 3A(s,t,u)+A(t,s,u)+A(u,t,s)\, , \nonumber \\
T^1(s,t,u)&=& A(t,s,u)-A(u,t,s)\, , \nonumber \\
T^2(s,t,u)&=& A(t,s,u)+A(u,t,s)\, .
\end{eqnarray}
Partial wave amplitudes, projected out via
\begin{equation}
T_l^I(s)={1\over 64\pi}\int^1_{-1} d(\cos\theta) P_l(\cos \theta)
T^I(s,t,u)  \, ,
\end{equation}
can be used to identify the associated scattering phase shifts via
\begin{equation}
 T^I_l(s) = \left( {s\over s-4m^2_{\pi}}\right)^{1\over 2} e^{i\delta^I_l}
\sin \delta^I_l \,  .
\end{equation}
Then from the lowest order chiral form
\begin{equation}
A(s,t,u)={s-m^2_{\pi}\over F^2_{\pi}}
\end{equation}
we determine values for
the pion scattering lengths and effective ranges
\begin{eqnarray}
a^0_0&=&{7m^2_{\pi}\over 32\pi F^2_{\pi}}\, , \quad
a^2_0=-{m^2_{\pi}\over 16\pi F^2_{\pi}}\, , \quad
a^1_1=-{m^2_{\pi}\over 24\pi F^2_{\pi}}\, ,  \nonumber \\
b^0_0&=&{m^2_{\pi}\over 4\pi F^2_{\pi}}\, , \quad
b^2_0={m^2_{\pi}\over 8\pi F^2_{\pi}}\, ,
\end{eqnarray}
comparison of which with experimental numbers is shown in Table 1.

\begin{table}
\begin{center}
\begin{tabular}{l r r r}
\hline \hline
& Experimental & Lowest Order\footnotemark[3]&
First Two Orders\footnotemark[3] \\ \hline
$a^0_0$ & $0.26\pm  0.05 $& 0.16 & 0.20 \\
$b^0_0$ &$ 0.25\pm  0.03$ & 0.18 & 0.26 \\
$a^2_0$ &$-0.028 \pm  0.012$ & -0.045 &-0.041 \\
$b^2_2$ &$-0.082 \pm  0.008$ & -0.089 &-0.070 \\
$a^1_1$ &$0.038 \pm  0.002$ & 0.030 &0.036 \\
$b^1_1$ -- & 0 & 0.043 \\
$a^0_2$ & $(17\pm 3) \times 10^{-4 }$& 0 & $20 \times 10^{-4}$ \\
$a^2_2$ & $(1.3\pm 3) \times 10^{-4 }$& 0 & $3.5 \times 10^{-4}$ \\
\hline\hline
\end{tabular}
\caption{The pion scattering lengths and slopes compared with
predictions of chiral symmetry.}
\end{center}
\end{table}

\par \bigskip

Despite the obvious success of this and other such predictions,\cite{gg} 
it is clear that we do not
really have at this point a satisfactory theory, since the strictures of unitarity
are violated.  Indeed, since we are working at tree level, all our amplitudes are
real.  However, unitarity of the S-matrix requires transition amplitudes to
contain an imaginary component since
\begin{eqnarray}
0&=&S^\dagger S-1=i(<f|T^\dagger|i> -<f|T|i>)+<f|T^\dagger T|i>\nonumber\\
&& i.e.\quad 
2{\rm Im}<f|T|i>=\sum_n <f|T^\dagger|n><n|T|i>\neq 0
\end{eqnarray}
The solution of such problems with unitarity are well known---the
inclusion of loop corrections to these simple tree level
calculations.  Insertion of such loop terms removes the unitarity violations
but comes with a high price---numerous divergences are introduced and this
difficulty prevented progress in this field for nearly a decade until a paper by
Weinberg suggested the solution.\cite{wbp}  One can deal with such divergences,
just as in QED, by introducing phenomenologically determined 
counterterms into the Lagrangian in order
to absorb the infinities.  We see in the next section how this can be accomplished.

\section{Renormalization}

\subsection{Effective Chiral Lagrangian}

We can now apply Weinberg's solution to the effective chiral Lagrangian, Eq. 
\ref{eq:abc}.  As noted above, when loop corrections are made to lowest order
amplitudes in order to enforce unitarity, divergences inevitably arise.
However, there is an important difference from the familiar case of QED in that the
form of the divergences is {\it different} from their lower order
counterparts---{\it i.e.} the theory is nonrenormalizable!  
The reason for this can be seen from a simple example.
Thus consider pi-pi scattering.  In lowest order there exists a
tree level contribution from ${\cal L}_2$ which is ${\cal O} (p^2/F_\pi^2)$
where $p$ represents some generic external energy-momentum.  The fact that
$p$ appears to the second power is due to the feature that its origin is
the {\it two}-derivative Lagrangian ${\cal L}_2$.  Now suppose that pi-pi
scattering is examined at one loop order.  Since the scattering amplitude
must still be dimensionless but now the amplitude involves a factor
$1/F_\pi^4$ the numerator must involve {\it four} powers of energy-momentum.
Thus any counterterm which is included in order to absorb this divergence
must be {\it four}-derivative in character.  Gasser and Leutwyler have studied
this problem and have written the most general form of such an order four
counterterm in chiral SU(3) as\cite{cpt}

\begin{eqnarray}
{\cal L}_4 &  =&\sum^{10}_{i=1} L_i {\cal O}_i
= L_1\bigg[{\rm tr}(D_{\mu}UD^{\mu}U^{\dagger})
\bigg]+L_2{\rm tr} (D_{\mu}UD_{\nu}U^{\dagger})\cdot
{\rm tr} (D^{\mu}UD^{\nu}U^{\dagger}) \nonumber \\
 &+&L_3{\rm tr} (D_{\mu}U D^{\mu}U^{\dagger}
D_{\nu}U D^{\nu}U^{\dagger})
+L_4 {\rm tr}  (D_{\mu}U D^{\mu}U^{\dagger})
{\rm tr} (\chi{U^{\dagger}}+U{\chi}^{\dagger}
) \nonumber \\
&+&L_5{\rm tr} \left(D_{\mu}U D^{\mu}U^{\dagger}
\left(\chi U^{\dagger}+U \chi^{\dagger}\right)
\right)+L_6\bigg[ {\rm tr} \left(\chi U^{\dagger}+
U \chi^{\dagger}\right)\bigg]^2 \nonumber \\
&+&L_7\bigg[ {\rm tr} \left(\chi^{\dagger}U-
U\chi^{\dagger}\right)\bigg]^2 +L_8 {\rm tr}
\left(\chi U^{\dagger}\chi U^{\dagger}
+U \chi^{\dagger}
U\chi^{\dagger}\right)\nonumber \\
&+&iL_9 {\rm tr} \left(F^L_{\mu\nu}D^{\mu}U D^{\nu}
U^{\dagger}+F^R_{\mu\nu}D^{\mu} U^{\dagger}
D^{\nu} U \right) +L_{10} {\rm tr}\left(F^L_{\mu\nu}
U F^{R\mu\nu}U^{\dagger}\right) \nonumber\\
\end{eqnarray}
where the covariant derivative is defined via
\begin{equation}
D_\mu U=\partial_\mu U+\{A_\mu,U\}+[V_\mu,U],
\end{equation}
the constants $L_i, i=1,2,\ldots 10$ are arbitrary (not determined from chiral
symmetry alone) and
$F^L_{\mu\nu}, F^R_{\mu\nu}$ are external field strength tensors defined via
\begin{eqnarray}
F^{L,R}_{\mu\nu}=\partial_\mu F^{L,R}_\nu-\partial_\nu
F^{L,R}_\mu-i[F^{L,R}_\mu ,F^{L,R}_\nu],\qquad F^{L,R}_\mu =V_\mu\pm A_\mu .
\end{eqnarray}
Now just as in the case of QED the bare parameters $L_i$ which appear
in this Lagrangian are not physical quantities.  Instead the experimentally 
relevant (renormalized)
values of these parameters are obtained by appending to these bare values
the divergent one-loop contributions having the form
\begin{equation} L^r_i = L_i -{\gamma_i\over 32\pi^2}
\left[{-2\over \epsilon} -\ln (4\pi)+\gamma -1\right]\end{equation}
By comparing with experiment, Gasser and Leutwyler were able to determine empirical
values for each of these ten parameters.  While ten sounds like a rather
large number, we shall see below that this picture is actually quite {\t predictive}.
Typical values for the parameters are shown in Table 2.
\begin{table}
\begin{center}
\begin{tabular}{l l c}\hline\hline
Coefficient & Value & Origin \\
\hline
$L_1^r$ & $0.65\pm 0.28$ & $\pi\pi$ scattering \\
$L_2^r$ & $1.89\pm 0.26$ & and\\
$L_3^r$ & $-3.06\pm 0.92$ & $K_{\ell 4}$ decay \\
$L_5^r$ & $2.3\pm 0.2$ & $F_K/F_\pi$\\
$L_9^r$ & $7.1\pm 0.3$ & $\pi$ charge radius \\
$L_{10}^r$ & $-5.6\pm 0.3$ & $\pi\rightarrow e\nu\gamma$\\
\hline\hline
\end{tabular}
\caption{Gasser-Leutwyler counterterms and the means by which
they are determined.}
\end{center}
\end{table}

The important question to ask at this point is why stop at order four derivativeas
Clearly if two loop amplitudes from ${\cal L}_2$ or one-loop
corrections from ${\cal L}_4$ are calculated, divergences will arise which
are of six derivative character.  Why not include these?  The answer is that
the chiral procedure represents an expansion in energy-momentum.  Corrections
to the lowest order (tree level) predictions from one loop corrections from ${\cal L}_2$
or tree level contributions from ${\cal L}_4$ are ${\cal
O}(E^2/\Lambda_\chi^2)$
where $\Lambda_\chi\sim 4\pi F_\pi\sim 1$ GeV is the chiral scale.\cite{sca}
Thus chiral
perturbation theory is a {\it low energy} procedure.  It is only to the extent
that the energy is small compared to the chiral scale that it makes sense to
truncate the expansion at the four-derivative level.  Realistically this
means that we deal with processes involving $E<500$ MeV, and, as we shall
describe below, for such reactions the procedure is found to work very well.

Now Gasser and Leutwyler, besides giving the form of the ${\cal O}(p^4)$ chiral
Lagrangian, have also performed the one loop integration and have written the
result in a simple algebraic form.  Users merely need to look up the result in
their paper.  However, in order to really understand what they have
done, it is useful to study a simple example of a chiral perturbation theory 
calculation
in order to see how it is performed and in order to understand how the experimental
counterterm values are actually determined.  We consider the pion
electromagnetic form factor, which by Lorentz- and gauge-invariance has the
structure
\begin{equation} \langle \pi^+(p_2)|J^{\mu}_{\mbox{em}}|\pi^+(p_1)
\rangle=
F_1(q^2)(p_1+p_2)^{\mu} \end{equation}
We begin by identifying the electromagnetic current as
\begin{eqnarray}
J^{\mu}_{\mbox{em}}& =& -{\partial{\cal L}\over \partial
(eA_{\mu})}=(\varphi \times \partial^{\mu}\varphi)_3\bigg[
1-{1\over 3F^2}\varphi\cdot \varphi+{\cal O}(\varphi^4)\bigg]
\nonumber \\
&+&(\varphi\times\partial^{\mu}\varphi)_3
\bigg[16L_4
+8L_5\bigg]{m^2_{\pi}\over F^2}+
{4L_9\over F^2}\partial^{\nu}(\partial^{\mu}
\varphi\times\partial_{\nu}\varphi)_3+
\cdots
\end{eqnarray}
where we have expanded to {\it fourth} order in the pseudoscalar fields.
Defining

\begin{eqnarray}
\delta_{jk}I(m^2)&=&i\Delta_{Fjk}(0)=\langle 0|\varphi_j(x)
\varphi_k(x)|0\rangle , \nonumber \\
I(m^2)&=&\mu^{4-d}\int{d^d k\over (2\pi)^d}{i\over
 k^2-m^2}={\mu^{4-d}\over (4\pi)^{d/2}} \Gamma
\left(1-{d\over 2}\right)(m^2)^{{d\over 2}-1} , \nonumber \\
\delta_{jk}I_{\mu\nu}(m^2) &=& -\partial_{\mu}\partial_{\nu}
i\Delta_{Fjk}(0)=\langle 0|\partial_{\mu}\varphi_j(x)\partial_{\nu}
\varphi_k(x)|0\rangle , \nonumber \\
I_{\mu\nu}(m^2) & =& \mu^{4-d} \int{d^d k\over (2\pi)^d }
k_{\mu}k_{\nu} {i\over k^2 -m^2}=g_{\mu\nu}{m^2\over d}
I(m^2)
\end{eqnarray}
we calculate the one loop correction shown in Figure 5a to be

\begin{equation} J^{\mu}_{\mbox{em}}|_{(4a)}= -{5\over
 3F^2_{\pi}}
(\varphi\times\partial^{\mu}\phi)_3I
(m^2_{\pi}) \end{equation}

\begin{figure}
\begin{center}
\epsfig{file=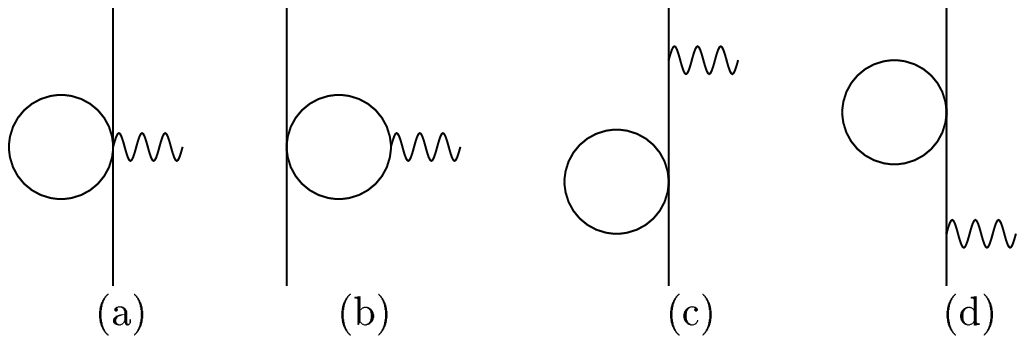}
\caption{Loop corrections to the pion form factor.}
\end{center}
\end{figure}

We also need the one loop correction shown in Figure 5b.  For this
piece we require the form of the pi-pi scattering amplitude which
arises from ${\cal L}_2$

\begin{equation}
\langle\pi^+(k_1)\pi^-(k_2)|\pi^+(p_1)\pi^-(p_2)\rangle
= {i\over 3F_0^2}\left(2m^2_0+p^2_1+p^2_2+
k^2_1+k^2_2-3(p_1-k_1)^2\right)
\end{equation}
and we shall perform the loop integration using the method of dimensional 
regularization, which yields
\begin{eqnarray}
\langle J^{\mu}_{\mbox{em}}\rangle _{(5b)} &=& {1\over (4\pi
F_{\pi})^2} (p_1+p_2)^{\mu} \int^1_0 dx(m^2_{\pi} -q^2
 x(1-x)) \nonumber \\
&\times& \bigg[\left(-{2\over \epsilon}+\gamma-1-\ln 4\pi\right)
+\ln {m^2_{\pi}-q^2 x(1-x)\over \mu^2 }\bigg]
\end{eqnarray}
Performing the x-integration we find, finally
\begin{eqnarray}
\langle J^{\mu}_{\mbox{em}}\rangle _{(5b)} &=& {1\over (4\pi
F_{\pi})^2} (p_1+p_2)^{\mu} \bigg\{\left(m^2_{\pi}-
{1\over 6} q^2 \right) \bigg[ -{2\over \epsilon} +\gamma -1
-\ln 4\pi +\ln {m^2_{\pi}\over \mu^2}\bigg] \nonumber  \\
&& + {1\over 6} (q^2-4m^2_{\pi})H\left({q^2\over
m^2_{\pi}}\right)-{1\over 18} q^2\bigg\},
\end{eqnarray}
where the function H(a) is given by
\begin{eqnarray}
H(a) & \equiv & \int_0^1 dx \ln (1-a x(1-x)) \nonumber \\
& = & \Bigg\{\begin{array}{l l}
2-2\sqrt{{4\over a}-1} \cot^{-1}\sqrt{{4\over a}-1}& (0<a<4) \\
2+\sqrt{1-{4\over a}}\bigg[\ln {\sqrt{1-{4\over a}}-1 \over
\sqrt{1-{4\over a}} +1} + i\pi \theta (a-4) \bigg] &
\mbox{(otherwise) }
\end{array}
\end{eqnarray}
and contains the imaginary component required by unitarity.

We are not done yet, however, since we must also include mass and
wavefunction effects---figs. 5c,5d.  In order to do so, we expand ${\cal L}_2$
to fourth order in $\varphi(x)$, and ${\cal L}_4$ to second order:
\begin{eqnarray}
{\cal L}_2&=&{1\over 2} [\partial^{\mu} \varphi\cdot
 \partial_{\mu} \varphi -m^2_0\varphi\cdot\varphi]+
{m^2_0\over 24 F^2_0}(\varphi\cdot\varphi)^2 \nonumber \\
&& +{1\over 6F^2_0}[(\varphi\cdot\partial^{\mu}
\varphi)(\varphi\cdot\partial_{\mu}\varphi)-(\varphi
\cdot\varphi)(\partial^{\mu}\varphi\cdot\partial_{\mu}
\varphi)]+{\cal O}(\varphi^6), \nonumber \\
{\cal L}_4&=&{m^2_0 \over F^2_0}[16L_4
+8L_5]{1\over 2}\partial_{\mu}
\varphi\cdot\partial^{\mu}\varphi \nonumber \\
&& -{m^2_0\over F^2_0}[32L_6+16
L_8]{1\over 2} m^2_0\varphi\cdot\varphi+
{\cal O}(\varphi^4) .
\end{eqnarray}
Performing the loop integrations on the $\phi^4(x)$ component of the above
yields
\begin{eqnarray}
{\cal L}_{\mbox{eff}}& =& {1\over 2} \partial^{\mu} \varphi
\partial_{\mu} \varphi -{1\over 2} m^2_0 \varphi \cdot
 \varphi +{5m^2_{\pi} \over 12F^2_{\pi}}
 I(m^2_{\pi})\varphi\cdot \varphi \nonumber \\
&&+{1\over 6F^2_{\pi}}(\delta_{ik}\delta_{jl}-\delta_{ij}
\delta_{kl}) I(m^2_{\pi})(\delta_{ij}\partial^{\mu}\varphi_k
\partial_{\mu} \varphi_l+\delta_{kl}m^2_{\pi}\varphi_i\varphi_j)
\nonumber  \\
&& +{1\over 2} \partial_{\mu}\varphi\cdot \partial^{\mu}
\varphi{m^2_{\pi}\over F^2_{\pi}}[16L_4+8
L_5)]-{1\over 2}m^2_{\pi}\varphi\cdot\varphi
{m^2_{\pi}\over F^2_{\pi}}[32L_6+16
L_8\bigg] \nonumber \\
&=&{1\over 2}\partial^{\mu}\varphi\cdot\partial_{\mu}
\varphi \bigg[1+(16L_4+8L_5)
{m^2_{\pi}\over F^2_{\pi}} -{2\over 3F^2_{\pi}}I(m^2_{\pi})
\bigg] \nonumber \\
&& -{1\over 2} m^2_0\varphi\cdot\varphi \bigg[
 1+(32L_6+16L_8){m^2_{\pi}\over
F^2_{\pi}} -{1\over 6F^2_{\pi}}I(m^2_{\pi})\bigg]
\end{eqnarray}
from which we can now read off the wavefunction renormalization
term $Z_\pi$.

When this is done we find

\begin{eqnarray}
Z_{\pi}F_1^{\mbox{(tree)}}(q^2)&=&
\bigg[1-{8m^2_{\pi}\over
 F^2_{\pi}}(2L_4 +L_5 \nonumber \\
& & +{m^2_{\pi}\over 24\pi^2 F^2_{\pi}}
\bigg\{-{2\over \epsilon}+\gamma -1-\ln
4\pi +\ln {m^2_{\pi}\over \mu^2}\bigg\}\bigg]
 \nonumber \\
& & \times\bigg[1+{8m^2_{\pi}\over F^2_{\pi}}
(2L_4+L_5)+2 q^2{L_9
\over F^2_{\pi}} \bigg] \nonumber  \\
&=& \bigg[ 1+{m^2_{\pi}\over 24\pi^2
 F^2_{\pi}} \bigg(-{2
\over \epsilon} +\gamma -1 -\ln 4\pi +\ln {m^2_{\pi}\over
 \pi^2}\bigg)+{2L_9\over F^2_{\pi}}q^2 \bigg]\nonumber\\
\end{eqnarray}
while from the loop diagrams given earlier
\begin{eqnarray}
F_1(q^2)\Bigg|_{(5a)}&=& -{5m^2_{\pi}\over 48\pi^2
F^2_{\pi}}\Bigg\{-{2\over \epsilon}+\gamma-1-\ln 4\pi
 +\ln{m^2_{\pi}\over \mu^2}\Bigg\} \nonumber \\
F_1(q^2)\Bigg|_{(5b)}&=& {1\over 16\pi^2 F^2_{\pi}}
\Bigg\{\Bigg(m^2_{\pi}-{1\over 6}q^2\Bigg)
\Bigg[ -{2\over \epsilon}
+\gamma -1 -\ln 4\pi +\ln {m^2_{\pi}\over \mu^2}\Bigg]
\nonumber \\
&& \left.
+{1\over 6}(q^2-4m^2_{\pi})H\left({q^2\over m^2_{\pi}}
\right)-{1\over 18}q^2\right\}
\end{eqnarray}
Adding everything together we have the final result, which when
written in terms of the renormalized value $L_9^{(r)}$ is {\it finite}!
\begin{eqnarray}
F_1(q^2) &=& 1+{2L_9^{(r)}\over
 F^2_{\pi}}q^2
+{1\over 96\pi^2F^2_{\pi}}
\left[(q^2-4m^2_{\pi})H\left({q^2\over
 m^2_{\pi}}\right)-q^2\ln {m^2_{\pi}\over
 \mu^2}-{q^2\over 3}\right] \nonumber\\
&&\label{eq:bh}
\end{eqnarray}
Expanding to lowest order in $q^2$ we find
\begin{eqnarray}
F_1(q^2) &=& 1+q^2\left[
{2L^{(r)}_9\over F^2_{\pi}} -{1\over 96
\pi^2F^2_{\pi}}\left(\ln {m^2_{\pi}\over \mu^2}+
1\right)\right]+\cdots
\end{eqnarray}
which can be compared with the phenomenological description in terms of
the pion charge radius
\begin{equation} F_1(q^2)=1+{1\over 6}\langle
 r^2_{\pi}\rangle q^2 +\cdots \end{equation}
By equating these two expressions and using the experimental value
of the pion charge radius---$\langle r_\pi^2\rangle_{\rm exp}=
(0.44\pm 0.01) \mbox{fm}^2$\cite{pdb}---we determine the value of the
counterterm $L_9^{(r)}$ shown in Table 2.

\begin{figure}
\begin{center}
\epsfig{file=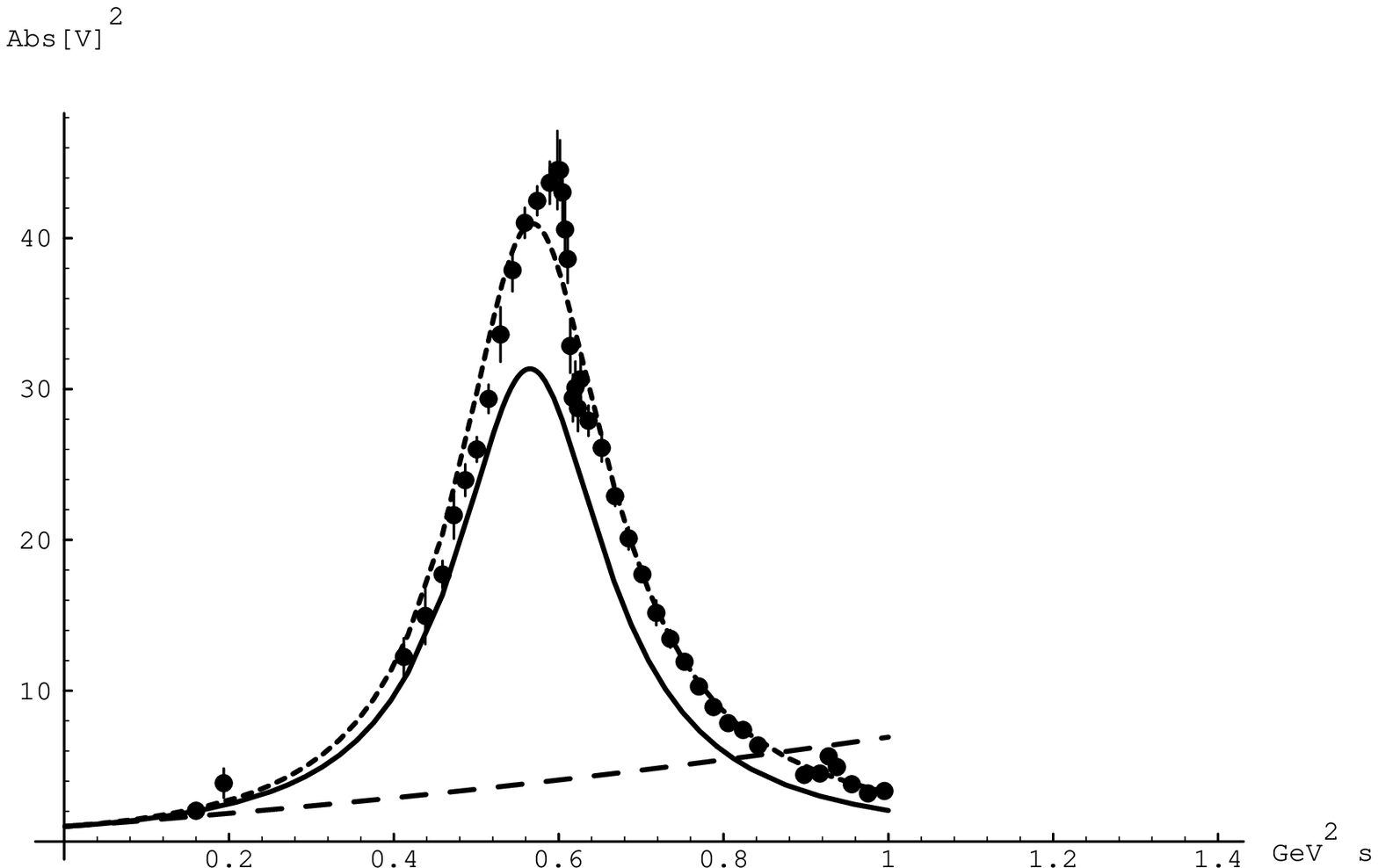,height=6cm,width=10cm}
\caption{Calculations of the modulus of the pion form factor squared 
compared to experimental results. Here
the solid line gives the result of the inverse amplitude method, while the dashed line
gives the one loop chiral perturbation theory prediction.  The dotted line shows an 
empirical simulation of the inelastic $omega\pi$ contribution obtained by multiplying
the inverse amplitude result by the factor $1+0.15s/s_\omega$.}
\end{center}
\end{figure}

As seen in Figure 6\cite{tru} this form gives a reasonable representation of the
experimental pion form factor near threshold but deviates substantially from
the empirical result as the $\rho$ resonance is approached.  This is not 
surprising as any perturbative approach will be unable to reproduce resonant
behavior.  This failure should not be considered a failure of chiral perturbative
techniques per se---just that as one approaches higher energy the importance of
two-loop (${\cal O}(p^6)$) and higher terms become important.  Although for
simple processes such two loop studies have been performed, the
number of $p^6$ counterterms is well over a hundred and a general chiral analysis
at two loop level is not feasible.  Nevertheless things are certainly not
hopeless, and in the closing chapter of these lectures we present some approaches
to extend the validity of chiral predictions to higher energy.

More relevant at this point is to stay near threshold and ask if 
chiral pertubation methods are predictive.  Can they be used as a {\it test} of QCD, for 
example?  The answer is definitely yes!  We do not have time in these lectures to give a
detailed presentation of the status of such tests--a simple example
will have to suffice.\cite{baj}  We have seen above how the 
pion charge radius enables the
determination of one of the chiral parameters---$L_9^r$.  A second---$L_{10}^r$ can
be found from measurement of the axial structure constant---$h_A$---in 
radiative pion decay---$\pi^+\rightarrow e^+\nu_e\gamma$ or $\pi^+\rightarrow
e^+\nu_e e^+e^-$, for which the decay amplitudes can be written 
\begin{eqnarray}
{\cal M}_{\pi^+\rightarrow e^+\nu_e\gamma}
&=&-{eG_F\over \sqrt{2}} \cos\theta_1 M_{\mu
\nu}(p,q)\varepsilon^{\mu\ast}(q)\bar{u}(p_{\nu})
\gamma^{\nu}(1+\gamma_5)v(p_e) \nonumber \\
{\cal M}_{\pi^+\rightarrow e^+\nu_e e^+ e^-}
&=&-{e^2G_F\over \sqrt{2}} \cos\theta_1 M_{\mu
\nu}(p,q){1\over q^2} \nonumber \\
&& \times \bar{u}(p_2)
\gamma^{\mu} v(p_1)\bar{u}(p_{\nu}) \gamma^{\nu}
 (1+\gamma_5)v(p_e),
\end{eqnarray}
ans the hadronic component of $M_{\mu\nu}$ has the structure
\begin{eqnarray}
M_{\mu\nu}(p,q)&=&\int d^4xe^{iq\cdot x}<0|T(J^{\rm em}_\mu (x)
J_\nu^{1-i2}(0)|\pi(\vec{p})>
= \mbox{Born terms}\nonumber\\
&-&h_A((p-q)_\mu q_\nu -g_{\mu\nu}q\cdot (p-q))
-r_A(q_\mu q_\nu -g_{\mu\nu}q^2)\nonumber\\
&+&ih_V\epsilon_{\mu\nu\alpha\beta}q^\alpha
p^\beta
\end{eqnarray}
where $h_A,r_A,h_V$ are unknown structure functions.  (Note that $r_A$
can be measured only via the rare Dalitz decay $\pi^+\rightarrow e^+
\nu_e e^+e^-$.) 

We also note that the related amplitude for Compton
scattering can be written in the form
\begin{eqnarray}
-iT_{\mu\nu}(p,p',q)&=&-i\int d^4x e^{iq_1\cdot x} <\pi^+(\vec{p}')|
T(J_\mu^{\rm em}(x)J_\nu^{\rm em}(0)|\pi^+(\vec{p})>\nonumber\\
&=&\mbox{Born terms}+\sigma(q_{2\mu}q_{1\nu}-g_{\mu\nu}q_1\cdot q_2)+\cdots
\end{eqnarray}
The $\gamma\pi^+\rightarrow\gamma\pi^+$ reaction is often analyzed in terms of
the pion electric and magnetic polarizabilities $\alpha_E$ and $\beta_M$ which
describe the response of the pion to external electric and magnetizing 
fields.\cite{brh}  In the
static limit such fields induce electric and magnetic dipole moments
\begin{equation}
\vec{p}=4\pi\alpha_E\vec{E},\quad \vec{\mu}=4\pi\beta_M\vec{H}
\end{equation}
which correspond to an interaction energy
\begin{equation}
U=-{1\over 2}\left(4\pi\alpha_E\vec{E}^2+4\pi\beta_M\vec{H}^2\right)
\end{equation}
Use of chiral perturbation theory yields the results
\begin{eqnarray}
h_V&=&{N_c\over 12\sqrt{2}\pi^2F_\pi}=0.027m_\pi^{-1},\quad {h_A\over h_V}=32\pi^2
(L_9^r+L_{10}^r)\nonumber\\
{r_A\over h_V}&=&32\pi^2\left[L_9^r-{1\over 192\pi^2}\left(\ln{m_\pi^2\over \mu^2}
+1\right)\right], \quad \alpha_E+\beta_M=0\nonumber\\
\alpha_E&=&{\alpha\over 2m_\pi}\sigma={4\alpha\over m_\pi F_\pi^2}\left(
L_9^r+L_{10}^r\right)
\end{eqnarray}
Use of the experimental result
\begin{equation}
{h_A\over h_V}=0.46\pm 0.08\quad{\rm gives}\quad L_{10}^r(\mu=m_\eta)=-0.0056(3)
\end{equation} 
and once this is determined chiral symmetry makes {\it four} predictions 
among these parameters!  As shown in Table 3, three of the four are found to be 
in good agreement with experiment.  The possible exception involves a relation
between the charged pion polarizability and the axial structure constant $h_A$
 measured in radiative pion decay.  In this case there exist three conflicting
experimental results, one of which agrees and one of which does not agree
with the theoretical prediction.  It is important to resolve this potential
discrepancy, since such chiral predictions are firm ones.  There is no
way (other than introducing perversely large higher order effects) to
bring things into agreement were some large violation of a chiral prediction
to be verified, since the only ingredient which goes into such predictions
is the (broken) chiral symmetry of QCD itself!

\begin{table}
\begin{center}
\begin{tabular}{cccc}\hline\hline
Reaction&Quantity&Theory&Experiment\\
\hline
$\pi^+\rightarrow e^+\nu_e\gamma$ & $h_V(m_\pi^{-1})$ & 0.027 
& $0.029\pm 0.017$\cite{pdg}\\
$\pi^+\rightarrow e^+\nu_ee^+e^-$ & $r_V/h_V$ & 2.6 & $2.3\pm 0.6$\cite{pdg}\\
$\gamma\pi^+\rightarrow\gamma\pi^+$ & $(\alpha_E+\beta_M)\,(10^{-4}\,{\rm fm}^3)$& 0
&$1.4\pm 3.1$\cite{anti}\\
      &$\alpha_E\,(10^{-4}\,{\rm fm}^3)$&2.8 & $6.8\pm 1.4$\cite{anti1}\\
 & & & $12\pm 20$\cite{russ}\\
 & & & $2.1\pm 1.1$\cite{slac}\\
\hline
\end{tabular}
\caption{Chiral Predictions and data in radiative pion processes.}
\end{center}
\end{table}

\section{Baryon Chiral Perturbation Theory}

Our discussion of chiral methods given above was limited to the study of
the interactions of the pseudoscalar mesons (would-be Goldstone bosons) 
with leptons and with each
other.  In the real world, of course, interactions with baryons also
take place and it is an important problem to develop a useful predictive
scheme based on chiral invariance for such processes.  Again much work
has been done in this regard,\cite{gss} but there remain important 
problems.\cite{bkm}  Writing
down the lowest order chiral Lagrangian at the SU(2) level is
straightforward---
\begin{equation}
{\cal L}_{\pi N}=\bar{N}(i\not\!\!{D}-m_N+{g_A\over 2}\rlap /{u}\gamma_5)N
\end{equation}
where $g_A$ is the usual nucleon axial coupling in the chiral limit, the
covariant derivative $D_\mu=\partial_\mu+\Gamma_\mu$ is given by
\begin{equation}
\Gamma_\mu={1\over 2}[u^\dagger,\partial_\mu u]-{i\over 2}u^\dagger
(V_\mu+A_\mu)u-{i\over 2}u(V_\mu-A_\mu)u^\dagger ,
\end{equation}
and $u_\mu$ represents the axial structure
\begin{equation}
u_\mu=iu^\dagger\nabla_\mu Uu^\dagger
\end{equation}
Expanding to lowest order we find
\begin{eqnarray}
{\cal L}_{\pi N}&=&\bar{N}(i\rlap /{\partial}-m_N)N+g_A
\bar{N}\gamma^\mu\gamma_5{1\over 2}\vec{\tau}N\cdot({i\over F_\pi}\partial_\mu\vec{\pi}
+2\vec{A}_\mu)\nonumber\\
&-&{1\over 4F_\pi^2}\bar{N}\gamma^\mu\vec{\tau}N\cdot\vec{\pi}\times
\partial_\mu\vec{\pi}+\ldots
\end{eqnarray}
which yields the Goldberger-Treiman relation,
connecting strong and axial couplings of the nucleon system\cite{gt}
\begin{equation}
F_\pi g_{\pi NN}=m_N g_A
\end{equation}
Using the present best values for these quantities, we find
\begin{equation}
92.4 \mbox{MeV}\times 13.05 =1206 \mbox{MeV}\quad\mbox{vs.}\quad 1189 \mbox{MeV}
= 939\mbox{MeV}\times 1.266
\end{equation}
and the agreement to better than two percent strongly confirms the validity
of chiral symmetry in the nucleon sector.  Actually the Goldberger--Treiman relation
is only strictly true at the unphysical point $g_{\pi NN}(q^2=0)$ and one expects
about a 1\% discrepancy to exist.  An interesting "wrinkle" in this regard
is the use of the so-called Dashen-Weinstein relation which uses simple SU(3)
symmetry breaking to predict this discrepancy in terms of corresponding numbers in the
strangeness changing sector.\cite{dw}
 
A second prediction of the lowest order chiral Lagrangian deals with charged pion
photoproduction.  As emphasized previously, chiral symmetry 
requires any pion coupling
to be in terms of a (co-variant) derivative.  Hence there exists a 
$\bar{N}N\pi^\pm\gamma$ contact interaction (the Kroll-Ruderman term)\cite{kr} which 
contributes to threshold charged pion photoproduction.  Here what is measured is 
the s-wave or $E_{0+}$ multipole, defined via
\begin{equation}
\mbox{Amp}=4\pi(1+\mu)E_{0+}\vec{\sigma}\cdot\hat{\epsilon}+\ldots
\end{equation}
where $\mu=m_\pi/M$.  In addition to the Kroll-Ruderman piece there exists, at the
two derivative level, a second contact term which arises from
\begin{equation}
{\cal L}^{(2)}_{\pi\gamma NN}={eg_A\over 8MF_\pi}v\cdot qP_+[(1+\tau_3)
\not\!\!{A}^\perp,\gamma_5\tau^a]P_+={eg_A\over 2MF_\pi}S\cdot\epsilon
v\cdot q(\tau^a+\delta^{a3})
\end{equation}
Adding these two contributions yields the result\cite{krth}
\begin{eqnarray}
E_{0+}&=&\pm{1\over 4\pi(1+\mu)}{eg_A\over \sqrt{2}F_\pi}(1\mp{\mu\over 2})=
{eg_A\over 4\sqrt{2}F_\pi}\left(\begin{array}{cc}
1-{3\over 2}\mu & \pi^+ \\
-1+{1\over 2}\mu & \pi^-
\end{array}\right)\nonumber\\
&=&\left\{\begin{array}{ll}
+26.3 \times 10^{-3}/m_\pi  & \pi^+n \\
-31.3 \times 10^{-3}/m_\pi  & \pi^-p
\end{array}\right.,
 \end{eqnarray}
and the numerical predictions are found to be in excellent agreement with the
present experimental results, as seen in the Table 4.
\begin{table}
\begin{center}
\begin{tabular}{cc}
$E_{0+}^{\pi^+n}$&$(+27.9\pm 0.5)\times 10^{-3}/m_\pi$\cite{95}\\
&$(+28.8\pm 0.7)\times 10^{-3}/m_\pi$\cite{96}    \\
&$(+27.6\pm 0.3)\times 10^{-3}/m_\pi$\cite{tri} \\
$E_{0+}^{\pi^-p}$&$(-31.4\pm 1.3)\times 10^{-3}/m_\pi$\cite{95} \\
&$(-32.2\pm 1.2)\times 10^{-3}/m_\pi$\cite{97} \\
&$(-31.5\pm 0.8)\times 10^{-3}/m_\pi$\cite{sal}
\end{tabular}
\caption{Experimental values for threshold pion photoproduction
multipoles.}
\end{center}
\end{table}

\subsection{Heavy Baryon Methods}

Extension to SU(3) gives additional successful predictions---the linear
Gell-Mann-Okubo relation as well as the generalized Goldbeger-Treiman
relation.  However, difficulties arise when one attempts to include
higher order corrections to this formalism.  The difference from the
Goldstone case is that there now exist {\it two} dimensionful
parameters---$m_N$ and $F_\pi$---in the problem rather than
{\it one}---$F_\pi$.  Thus loop effects can be of order $(m_N/4\pi F_\pi)^2
\sim 1$ and we no longer have a reliable perturbative scheme.  A
consistent power counting mechanism can be constructed provided that we
eliminate the nucleon mass from the leading order Lagrangian.  This is done by
considering the nucleon to be very heavy.  Then we can write its
four-momentum as\cite{jm}
\begin{equation}
p_\mu=Mv_\mu+k_\mu
\end{equation}
where $v_\mu$ is the four-velocity and satisfies $v^2=1$, while $k_\mu$
is a small off-shell momentum, with $v\cdot k<< M$.  One can construct
eigenstates of the projection operators $P_\pm = {1\over 2}(1\pm
\rlap /{v})$, which in the rest frame project out upper, lower
components of the Dirac wavefunction, so that\cite{bkkm}
\begin{equation}
\psi=e^{-iMv\cdot x}(H_v+h_v)
\end{equation}
where
\begin{equation}
H_v=P_+\psi,\qquad h_v=P_-\psi
\end{equation}
The effective Lagrangian can then be written in terms of $N,h$ as
\begin{equation}
{\cal L}_{\pi N}=\bar{H}_v{\cal A}H_v+\bar{h}_v{\cal B}H_v+
\bar{H}_v\gamma_0{\cal B}^\dagger\gamma_0h_v-\bar{h}_v{\cal C}h_v
\end{equation}
where the operators ${\cal A}, {\cal B},{\cal C}$ have the low energy
expansions
\begin{eqnarray}
{\cal A}&=&iv\cdot D+g_A u\cdot S +\ldots\nonumber\\
{\cal B}&=&i\not\!\!{D}^\perp-{1\over 2}g_A v\cdot u\gamma_5+\ldots\nonumber\\
{\cal C}&=&2M+iv\cdot D+g_A u\cdot S+\ldots
\end{eqnarray}
Here $D_\mu^\perp=(g_{\mu\nu}-v_\mu v_\nu)D^\nu$ is the transverse component
of the covariant derivative and $S_\mu={i\over 2}\gamma_5
\sigma_{\mu\nu}v^\nu$ is the Pauli-Lubanski spin vector and satisfies
the relations
\begin{equation}
S\cdot v=0,\quad S^2=-{3\over 4},\quad\{S_\mu,S_\nu\}={1\over 2}(v_\mu v_\nu-
g_{\mu\nu}),\quad [S_\mu,S_\nu]=i\epsilon_{\mu\nu\alpha\beta}v^\alpha S^\beta
\end{equation}
We see that the two components H,h are coupled in this expression for the
effective action.  However, the system may be diagonalized by use of 
the field transformation
\begin{equation}
h'=h-{\cal C}^{-1}{\cal B}H
\end{equation}
in which case the Langrangian becomes
\begin{equation}
{\cal L}_{\pi N}=\bar{H}_v({\cal A}+(\gamma_0{\cal B}^\dagger
\gamma_0){\cal C}^{-1}{\cal B})H_v-\bar{h}'_v{\cal C}h'_v
\end{equation}
The piece of the Lagrangian involving $H$ only contains the mass in the operator
${\cal C}^{-1}$ and is the effective Lagrangian that we desire.  The remaining
piece involving $h'_v$ can be thrown away, as it does not couple to the
$H_v$ physics.  (In path integral language we simply integrate out this
component yielding an uninteresting overall constant.)
Of course, when loops are calculated a set of
counterterms will be required and these are given at leading (two-derivative)
order by
\begin{eqnarray}
{\cal A}^{(2)}&=&{M\over F_\pi^2}(c_1\mbox{Tr}\chi_+
+c_2(v\cdot u)^2+c_3u\cdot u+c_4[S^\mu,s^\nu]u_\mu u_\nu\nonumber\\
&+&c_5(\chi_+-\mbox{Tr}\chi_+)
-{i\over
4M}[S^\mu,S^\nu]((1+c_6)F^+_{\mu\nu}+c_7\mbox{Tr}f^+_{\mu\nu}))\nonumber\\
{\cal B}^{(2)}&=&{M\over F_\pi^2}((-{c_2\over 4}i[u^\mu,u^\nu]+c_6f_+^{\mu\nu}
+c_7Trf_+^{\mu\nu})\sigma_{\mu\nu}\nonumber\\
&-&{c_4\over 2}v_\mu\gamma_\nu Tru^\mu u^\nu)\nonumber\\
{\cal C}^{(2)}&=&-{M\over F_\pi^2}(c_1Tr\chi_++(-{c_2\over 4}i[u^\mu,u^\nu]
+c_6f_+^{\mu\nu}+c_7trF_+^{\mu\nu})\sigma_{\mu\nu}\nonumber\\
&-&{c_3\over 4}Tr u^\mu u_\nu
-({c_4\over 2}+Mc_5)v_\mu v_\nu Tru^\mu u^\nu)
\end{eqnarray}
Expanding ${\cal C}^{-1}$ and the other terms in terms of
a power series in $1/M$ leads to an
effective heavy nucleon Lagrangian of the form (to ${\cal O}(q^3)$)
\begin{eqnarray}
{\cal L}_{\pi N}&=&\bar{H}_v\{{\cal A}^{(1)}+{\cal A}^{(2)}+{\cal A}^{(3)}
+(\gamma_0{\cal B}^{(1)\dagger}\gamma_0){1\over 2M}{\cal B}^{(1)}\nonumber\\
&+&{(\gamma_0{\cal B}^{(1)\dagger}\gamma_0){\cal B}^{(2)}+(\gamma_0{\cal B}
^{(2)\dagger}\gamma_0){\cal B}^{(1)}\over 2M}\nonumber\\
&-&(\gamma_0{\cal B}^{(1)\dagger}\gamma_0){i(v\cdot D)+g_A(u\cdot S)\over
(2M)^2}{\cal B}^{(1)}\}H_v+{\cal O}(q^4)\label{eq:def}
\end{eqnarray}
A set of Feynman rules can now be written down and a consistent power counting
scheme developed, as shown by Meissner and his collaborators.\cite{bkm}

\subsection{Applications}
As an example of the use of this formalism, called heavy baryon chiral perturbation
theory (HB$\chi$pt) consider the nucleon-photon interaction.  To lowest (one
derivative) order we have from ${\cal A}^{(1)}$
\begin{equation}
{\cal L}_{\gamma NN}^{(1)}=ie\bar{N}{1\over2}(1+\tau_3)\epsilon\cdot vN
\end{equation}
while at two-derivative level we find
\begin{eqnarray}
 {\cal L}_{\gamma NN}^{(2)}=\bar{N}\left\{{e\over
4M}(1+\tau_3)\epsilon\cdot(p_1+p_2)
+{ie\over 2M}[S\cdot \epsilon,S\cdot k](1+\kappa_S+\tau_3(1+\kappa_V)\right\}N
\nonumber\\
\quad
\end{eqnarray}
where we have made the identifications
$c_6=\kappa_V,\quad c_7={1\over 2}(\kappa_S-\kappa_V)$.  We can now reproduce
the low energy theorems for Compton scattering.  Consider the case of the
proton. At the two derivative level,
we have the tree level prediction 
\begin{equation}
(\gamma_0{\cal B}^{(1)\dagger}\gamma_0){1\over 2M}
{\cal B}^{(1)}|_{\gamma pp}={e^2\over 2M}\vec{A}_\perp^2
\end{equation}
which yields the familiar Thomson amplitude
\begin{equation}
\mbox{Amp}_{\gamma pp}=-{e^2\over M}\hat{\epsilon}'\cdot\hat{\epsilon}
\end{equation}
On the other hand at order $q^3$ we find a contribution from Born diagrams
with two-derivative terms at each vertex, yielding
\begin{eqnarray}
\mbox{Amp}_{\gamma pp}&=&({e\over M}^2){1\over \omega}
\bar{p}[(\hat{\epsilon}'
\cdot\vec{k}\vec{S}\cdot\hat{\epsilon}\times\vec{k}
-\hat{\epsilon}\cdot\vec{k}'\vec{S}\cdot
\hat{\epsilon}'\times\vec{k}')(1+\kappa_p)\nonumber\\
&+&i\vec{S}\cdot(\hat{\epsilon}\times\vec{k})\times
(\hat{\epsilon}'\times\vec{k}')(1+\kappa_p)^2]
\end{eqnarray}
The full result must also include contact terms at order $q^3$ from the last
piece of Eq. \ref{eq:def}
\begin{equation}
-eP_+\not\!\!{A}^\perp {iv\cdot D\over (2M)^2}e\not\!\!{A}^\perp P_+=
-{e^2\over 2M^2}\vec{S}\cdot\vec{A}\times\dot{\vec{A}}
\end{equation}
and from the third
\begin{equation}
{1\over 2M}P_+\left\{e\not\!\!{A}^\perp,\kappa_p\sigma_{\mu\nu}F^{\mu\nu}\right\}
P_+=\kappa_p{e^2\over M^2}\vec{S}\cdot\vec{A}\times\dot{\vec{A}}
\end{equation}
When added to the Born contributions the result can be expressed in the general
form\cite{bkm}
\begin{eqnarray}
\mbox{Amp}&=&\hat{\epsilon}\cdot\hat{\epsilon}'A_1+\hat{\epsilon}'\cdot
\vec{k}\hat{\epsilon}\cdot\vec{k}'A_2+i\vec{\sigma}\cdot(\hat{\epsilon}'
\times\hat{\epsilon})A_3\nonumber\\
&+&i\vec{\sigma}\cdot(\vec{k}'\times\vec{k})\hat{\epsilon}'\cdot\hat{\epsilon}
A_4+i\vec{\sigma}\cdot[(\hat{\epsilon}'\times\vec{k})\hat{\epsilon}\cdot\vec{k}'
-(\hat{\epsilon}\times\vec{k}')\hat{\epsilon}'\cdot\vec{k}]A_5\nonumber\\
&+&i\vec{\sigma}\cdot[(\hat{\epsilon}'\times\vec{k}')\hat{\epsilon}\cdot\vec{k}'
-(\hat{\epsilon}\times\vec{k})\hat{\epsilon}'\cdot\vec{k}]A_6\nonumber\\
\end{eqnarray}
with\footnote{Here we
have used the identity
$$\vec{\sigma}\cdot(\hat{\epsilon}'\times\vec{k}')\times(\hat{\epsilon}\times
\vec{k})=\vec{\sigma}\cdot(\vec{k}'\times\vec{k})\hat{\epsilon}\cdot
\hat{\epsilon}'
+\vec{\sigma}\cdot(\hat{\epsilon}'\times\hat{\epsilon})\vec{k}'\cdot\vec{k}
+\vec{\sigma}\cdot(\hat{\epsilon}\times\vec{k}')\hat{\epsilon}'\cdot\vec{k}
-\vec{\sigma}\cdot(\hat{\epsilon}'\times\vec{k})\hat{\epsilon}\cdot\vec{k}'
$$}
\begin{eqnarray}
A_1&=&-{e^2\over M},\quad A_2={1\over M^2\omega},\quad A_3={e^2\omega\over
2M^2}
(1+2\kappa-(1+\kappa)^2\hat{k}\cdot\hat{k}')\nonumber\\
A_4&=&-A_5=-{e^2(1+\kappa)^2\over 2M^2\omega},\quad A_6=-{e^2(1+\kappa)
\over 2M^2\omega}
\end{eqnarray}
which agrees with the usual result derived in this order via
Low's theorem.\cite{low}

\begin{figure}
\begin{center}
\epsfig{file=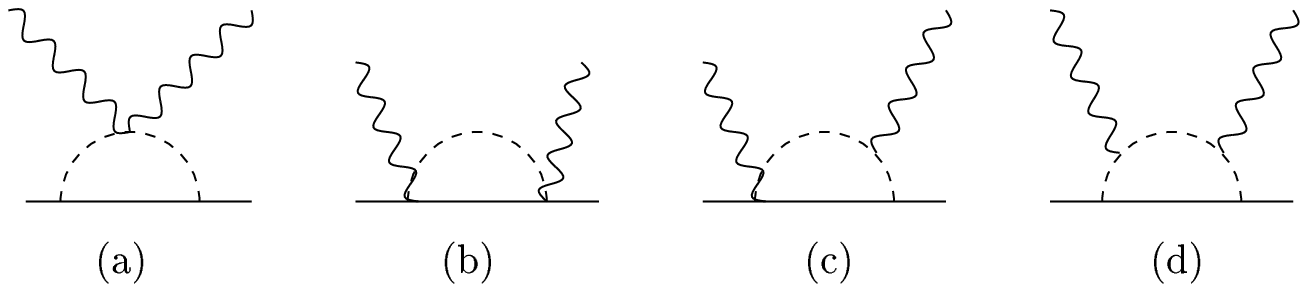}
\caption{Loop diagrams for Compton scattering.  Figures b,c,d must, of course, 
be augmented by appropriate cross diagrams.}
\end{center}
\end{figure}

A full calculation at order $q^3$ must also, of course,
include loop contributions.  Using the lowest
order (one-derivative) pion-nucleon interactions
\begin{eqnarray}
{\cal L}_{\pi NN}&=&{g_A\over F_\pi}\bar{N}\tau^aS\cdot q N\nonumber\\
{\cal L}_{\pi\pi NN}&=&{1\over 4F_\pi^2}v\cdot (q_1+q_2)\epsilon^{abc}\bar{N}
\tau_cN\nonumber\\
{\cal L}_{\gamma\pi NN}&=&{ieg_A\over F_\pi}\epsilon^{a3b}\bar{N}\epsilon\cdot
S\tau_bN
\end{eqnarray}
these can be calculated using the diagrams shown in Figure 7.  Of course, from
Eq. \ref{eq:def} the propagator for the nucleon must have the form $1/iv\cdot k$ where
k is the off-shell momentum.  Thus, for example, the seagull diagram, Figure
7a, is of the form
\begin{equation}
\mbox{Amp}=4e^2({g_A\over F_\pi})^2\hat{\epsilon}\cdot\hat{\epsilon}'
\int{d^4k\over (2\pi)^4}{S\cdot kS\cdot k\over v\cdot k(k^2-m_\pi^2)
((k+q_1-q_2)^2-m_\pi^2)}
\end{equation}
Since there are no additional counterterms at this order $q^3$, the sum of loop
diagrams must be finite and yields, to lowest order in energy and after
considerable calculation
\begin{eqnarray}
A_1^{loop}&=&\xi({11\omega^2\over 24m_\pi}+{t\over 48m_\pi}),\quad A_2^{loop}=
\xi({1\over 24m_\pi}),\quad A_3^{loop}=\xi({\omega t\over \pi m_\pi^2}+
{\omega^3\over 3\pi m_\pi^2})\nonumber\\
A_4^{loop}&=&\xi({\omega\over 6\pi m_\pi^2}),\quad A_5^{loop}=-A_6^{loop}=
-\xi({13\omega\over 12\pi m_\pi^2}),
\end{eqnarray}
and $\xi=g_A^2/8\pi F_\pi^2.$

The experimental implications of these results may be seen by first considering
the case of an unpolarized proton target.  Writing
\begin{equation}
\mbox{Amp}_{unpol}=\{\hat{\epsilon}\cdot\hat{\epsilon}'(-{e^2\over M}+4\pi
\alpha_E\omega^2)+(\hat{\epsilon}\times\vec{k})\cdot(\hat{\epsilon}'\times
\vec{k}')4\pi\beta_M\}
\end{equation}
where $\alpha_E,\beta_M$ are the proton electric and magnetic polarizabilities,
we identify the one loop chiral predictions\cite{bkmp}
\begin{equation}
\alpha_E^{theo}=10\beta_M^{theo}={5e^2g_A^2\over 384\pi^2F_\pi^2m_\pi}=
12.2\times 10^{-4}\mbox{fm}^3
\end{equation}
which are in reasonable agreement with the recently measured values\cite{nat}
\begin{equation}
\alpha_E^{exp}=(12.1\pm 0.8\pm 0.5)\times 10^{-4}\mbox{fm}^3,\qquad
\beta_M^{exp}=(2.1\mp 0.8\mp 0.5)\times
10^{-4}\mbox{fm}^3
\end{equation}
For the case of spin-dependent forward scattering, we find, in general
\begin{equation}
{1\over 4\pi}\mbox{Amp}=f_1(\omega^2)\hat{\epsilon}\cdot\hat{\epsilon}'
+i\omega f_2(\omega^2)\vec{\sigma}\cdot\hat{\epsilon}'\times\hat{\epsilon}
\end{equation}
with
\begin{eqnarray}
f_1(\omega^2)&=&-{e^2\over 4\pi M}+(\alpha_E+\beta_M)\omega^2+{\cal
O}(\omega^4)
\nonumber\\
f_2(\omega^2)&=&-{e^2\kappa_p^2\over 8\pi^2M^2}+\gamma_S\omega^2+{\cal
O}(\omega^4)
\end{eqnarray}
where $\gamma_S$ is a sort of "spin-polarizability" and is related to the classical
Faraday effect. Assuming that the amplitudes $f_1,f_2$ obey once-subtracted and 
unsubtracted dispersion relations respectively we find the sum rules
\begin{eqnarray}
\alpha_E+\beta_M&=&{1\over 4\pi^2}\int_{\omega_0}^\infty{d\omega\over \omega^2}
(\sigma_+(\omega)+\sigma_-(\omega))\nonumber\\
{\pi e^2\kappa_p^2\over 2M^2}&=&\int_{\omega_0}^\infty{d\omega\over \omega}
[\sigma_+(\omega)-\sigma_-(\omega)]\nonumber\\
\gamma_S &=& {1\over 4\pi^2}\int_{\omega_0}^\infty{d\omega\over \omega^3}
[\sigma_+(\omega)-\sigma_-(\omega)]
\end{eqnarray}
where here $\sigma_\pm(\omega)$ denote the photoabsorption cross sections
for scattering circularly polarized photons on polarized nucleons invovling total
$\gamma N$ helicity 3/2 and 1/2 respectively.
Here the first is the well-known Baldin sum rule for the sum of the electric
and magnetic polarizabilities, while the second is the equally familiar
Drell-Hearn-Gerasimov sum rule.\cite{dgh}  The third is less well known, but
follows
from that of DHG and offeres a new check of the chiral predictions.

Another venue where there has been a great deal of work is that of neutral
pion photoproduction, for which the
one-derivative contribution vanishes.  In this case the leading contribution
arises from the two derivative term given in Eq. \ref{eq:def}, augmented by the
three derivative contribution from Born diagrams.  The net result is
\begin{eqnarray}
\mbox{Amp}^{(2)}={eg_A\over 2F_\pi}\mu\hat{\epsilon}\cdot\vec{\sigma}\times
\left\{\begin{array}{cc}
1 & \pi^0p \\
0 & \pi^0n
\end{array}\right.
\end{eqnarray}
for the contact term and
\begin{equation}
\mbox{Amp}^{(3)}=-{e\over 2M}[S\cdot\epsilon,S\cdot k](1+\kappa_p)
{1\over v\cdot q}
{g_A\over 2MF_\pi}S\cdot(2p-q)m_\pi=
-{eg_A\over 4F_\pi}\mu^2(1+\kappa_p)\vec{\sigma}\cdot
\vec{\epsilon}
\end{equation}
for the pole terms (Note: only the cross term is nonvanishing at threshhold.)

\begin{figure}
\begin{center}
\epsfig{file=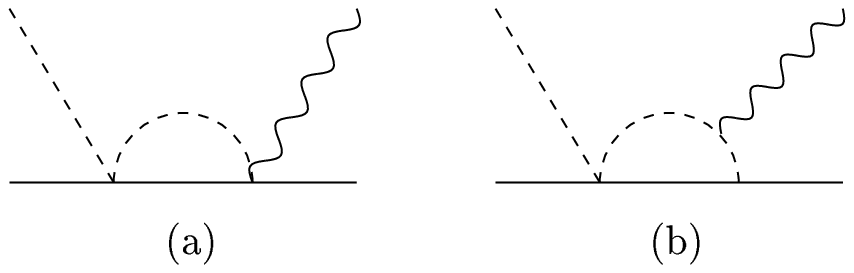}
\caption{Diagrams for neutral pion photoproduction.  Each should be
accompanied by an appropriate cross term.}
\end{center}
\end{figure}

Finally we must append the contribution of the loop contributions which arise
from the graphs shown in Figure 8
\begin{equation}
\mbox{Amp}^{loop}=-{eg_AM\over 64\pi
F_\pi^2}\mu^2\vec{\sigma}\cdot\vec{\epsilon}
\end{equation}
The result is the prediction\cite{vgkm}
\begin{equation}
E_{0+}={eg_A\over 8\pi M}\mu\{1-[{1\over 2}(3+\kappa_p)+({M\over 4F_\pi})^2]\mu
+{\cal O}(\mu^2)\}
\end{equation}
However, comparison with experiment is tricky because of the existence of
isotopic spin breaking in the pion and nucleon masses, so that
there are {\it two} thresholds---one for $\pi^0p$ and the second for
$\pi^+n$---only
7 MeV apart.  When the physical masses of the pions are used recent data 
from both Mainz and from Saskatoon agree with the
chiral prediction.  However, there are concerns about the convergence of the
chiral expansion, which reads $E_{0+}=C(1-1.26+0.59+\ldots)$. 
There also exist chiral predictions for threshold
p-wave amplitudes which are in good agreement with experiment, as shown in Table 5, and 
for which the convergence is calculated to be rapid.

\begin{table}
\begin{center}
\begin{tabular}{ccc}
 & theory&expt. \\
$E_{0+}(\pi^0p)(\times 10^{-3}/m_\pi)$& -1.2&$ -1.31\pm 0.08$\cite{mai}\\
 & & $-1.32\pm 0.11$\cite{salp}\\
$E_{0+}(\pi^0n)(\times 10^{-3}/m_\pi)$& 2.1& $1.9\pm 0.3$\cite{arg}\\
$P_1/|\vec{q}|(\pi p)(\times{\rm GeV}^{-2})$&0.48&$0.47\pm 0.01$\cite{mai}\\
 & &$0.41\pm 0.03$\cite{salp}
\end{tabular}
\caption{Threshold parameters for neutral pion photoproduction.}
\end{center}
\end{table}

Finally exists a chiral symmetry prediction for the reaction $\gamma n
\rightarrow \pi^0n$
\begin{equation}
E_{0+}=-{eg_A\over 8\pi M}\mu^2\{{1\over 2}\kappa_n+({M\over 4F_\pi})^2\}+\ldots=
2.13\times 10^{-3}/m_\pi
\end{equation}
However, the experimental measurement of such an amplitude involves
considerable challenge, and must be accomplished either by use of a deuterium
target with the difficult subtraction of the proton contribution and of
meson exchange contributions or by use of a ${}^3$He target.  Neither of these 
are straightforward although some limited data already exist.\cite{arg}

Other areas wherein chiral predictions can be confronted with experiment
include the electric dipole amplitude in electroproduction as well as in weak
interactions such as muon capture.  However, we do not have the space here to
cover this work.  We can succinctly summarize the situation by stating 
that at the present time there exist no significant
disagreements with experimental findings, but the predictive power in the
baryon sector is only in the near threshold region, due among other things 
to the feature that the expansion is in terms of $p/m$ rather than $p^2/m^2$
which occurs in the meson case. 

\section{On to Higher Energy: Dispersion Relations}

Above we have seen the power of chiral perturbation theory in addressing near-threshold
phenomenology.  However, we have also observed
its associated weakness---loss of predictive
ability in the meson sector once $E,p\geq \sim m_\rho/2$ and in the baryon-meson
sector even sooner.  In this closing chapter, we try to address the question of whether
one can somehow extend the success of $\chi$pt to higher energy without losing its
model-independent connections to QCD.  Strictly speaking the answer is no---as the
energy-momentum increases one {\it must} go to higher and higher order in the chiral 
expansion, which means increasing the number of loops and of unknown counterterms.
Any other approach must be model dependent at some level and hence use more than 
simply the (broken)
chiral symmetry of QCD as an input.  Nevertheless, there are some techniques
which keep such model dependence at a minimum.  One such procedure is to marry
the results of $\chi$pt with the strictures of dispersion relations,\cite{jfd} 
the validity
of which, given that they rely only on the causality properties of the theory, 
are not in question.  In the case of the pion form factor studied above in one loop
$\chi$pt the causality condition asserts that the function $F(q^2)$ is analytic
in the entire complex $q^2$-plane except for a cut along the real axis which 
extends from $4m_\pi^2<q^2<\infty$.  Along this cut $F(q^2)$ has an imaginary 
part which is given by the unitarity condition
\begin{eqnarray}
(p_2-p_1)_\mu{\rm Im}F_1(q^2)&=&{1\over 2}\int{d^3k_1\over (2\pi)^32E_1}
{d^3k_2\over (2\pi)^32E_2}(2\pi)^4\delta^4(p_1+p_2-k_1-k_2)\nonumber\\
&\times&<\pi\pi|T|\pi\pi>^*
<\pi\pi|j_\mu|0>\label{eq:pp}
\end{eqnarray}

\begin{figure}
\begin{center}
\epsfig{file=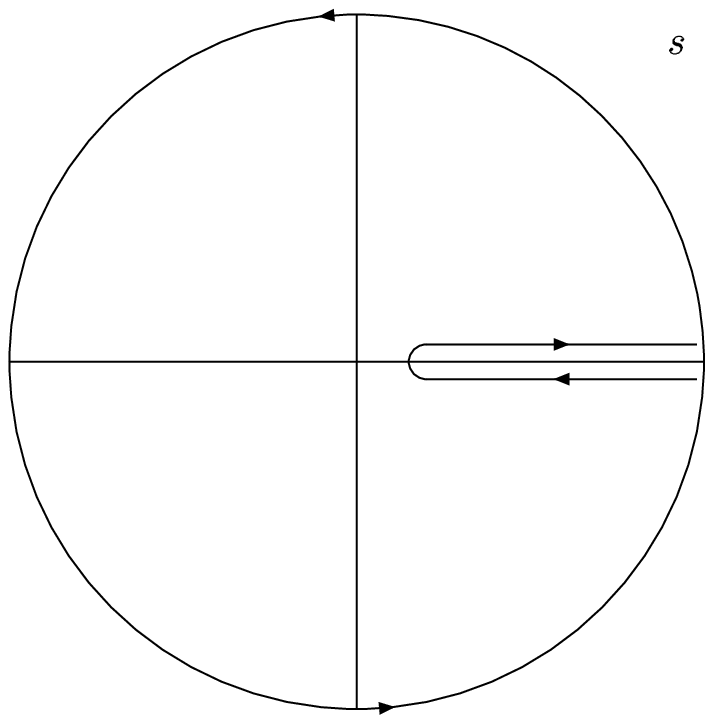}
\caption{Contour for form factor integration.}
\end{center}
\end{figure}

Use of Cauchy's theorem involving the contour as shown in Figure 9 and the assumption 
that $F(s)\rightarrow 0$ as $s\rightarrow \infty$ then leads to the dispersion relation
\begin{equation}
F_1(q^2)={1\over \pi}\int_{4m_\pi^2}^\infty {{\rm Im}F_1(s)ds\over s-q^2-i\epsilon}
\end{equation}    
If the asymptotic condition is {\it not} satisfied one can use various
numbers of subtractions in order to guarantee convergence.  For example, provided
that $F_1(s)/s\rightarrow 0$ as $s\rightarrow\infty$ we have
\begin{eqnarray}
F_1(q^2)-F_1(0)&=&{1\over \pi}\int_{4m_\pi^2}^\infty{\rm Im}F_1(s)ds\left(
{1\over s-q^2-i\epsilon}-{1\over s-i\epsilon}\right)\nonumber\\
&=&{q^2\over \pi}\int_{4m_\pi^2}^\infty
{{\rm Im}F_1(s)ds\over s(s-q^2-i\epsilon)}
\end{eqnarray}
and further subtractions may be performed if necessary.  Even if such subtractions
are not required it may be advantageous to utilize them.  The point is that in order
to get a prediction one requires the value of the imaginary component of the form
factor at {\it all} values of $s$.  However, the more subtractions the
less the sensitivity to input from large $s$, where in general the form factor is 
not as well determined.

In the case of the pion form factor it is convenient to perform two subtractions, whereby
\begin{equation}
F_1(q^2)=F_1(0)+q^2F_1'(0)+{q^4\over \pi}\int_{4m_\pi^2}^\infty
{{\rm Im}F_1(s)ds\over s^2(s-q^2-i\epsilon)}\label{eq:oo}
\end{equation}   
Here by current conservation we may set $F_1(0)=1$ while from experiment we have
$F_1'(0)=<r_\pi^2>/6=(0.073\pm 0.002)$ fm$^2$.  In order to find Im $F(s)$ we begin by
projecting onto the p-wave $\pi\pi$ channel via
\begin{equation}
T_1^1(s)={1\over 64\pi}\int_{-1}^1d(\cos\theta)P_1(\cos\theta)<\pi\pi|T|\pi\pi>
=\sqrt{s\over s-4m_\pi^2}e^{i\delta_1^1}\sin\delta_1^1
\end{equation}
whereby the unitarity condtion Eq. \ref{eq:pp} reads 
\begin{equation}
{\rm Im}F_1(s)=e^{-i\delta_1^1}\sin\delta_1^1F_1(s)\label{eq:rr}
\end{equation}
Of course, the solution of this equation is simply the Fermi-Watson theorem
\begin{equation}
F_1(s)=|F_1(s)|\exp i\delta_1^1(s)
\end{equation}
which states that the phase of the form factor is the p-wave $\pi\pi$ phase shift.
Now at lowest order we have the chiral prediction
\begin{equation}
T_1^1(s)={s-4m_\pi^2\over 96\pi F_\pi^2}\label{eq:tt}
\end{equation}
and if this is substituted into Eq. \ref{eq:rr} along with the lowest order result
$F_1(s)\approx F_1(0)=1$ we find
\begin{equation}
F_1(q^2)=1+{<r_\pi^2>\over 6}q^2+{1\over 96\pi^2F_\pi^2}\left((q^2-4m_\pi^2)H(q^2)
+{2\over 3}q^2\right)\label{eq:ss}
\end{equation} 
where we have used the integral
\begin{equation}
\int_{4m_\pi^2}^\infty{ds\over s^2}\sqrt{s-4m_\pi^2\over s}
\left(a+bs\over s-q^2-i\epsilon\right)={a+bq^2\over q^4}H(q^2)-
{a\over 6m^2q^2}\label{eq:uu}
\end{equation}
Comparison with the chiral perturbative result---Eq. \ref{eq:bh}---reveals 
that the results are identical provided that one identifies
\begin{equation}
{<r_\pi^2>\over 6}={2L_9^r\over F_\pi^2}+{1\over 96\pi^2F_\pi^2}
(1+\ln{m_\pi^2\over \mu^2})
\end{equation}
In this form then we see that the chiral perturbative predictions are simply a result
of use of the lowest order chiral forms in the input to the dispersion integral, while
a real world calculation would utilize experimental data.  In this context it is clear
that the renormalized counterterms simply play the role of subtraction constants.

So far our dispersive calculation has simply reproduced the chiral perturbative results.
How can one do better? Clearly by using experimental data rather then the simple lowest
order chiral forms.\cite{dok}
For example, if we divide Eq. \ref{eq:oo} by $q^2$ and then take the limit as
$q^2\rightarrow\infty$ we find a sum rule for $L_9^r(\mu)$
\begin{equation}
L_9^r(\mu)={F_\pi^2\over 6\pi}\int_{4m_\pi^2}^\infty{ds\over s^2}{\rm Im}F_1(s)
-{1\over 192\pi^2}(1+\ln{m_\pi^2\over \mu^2})\label{eq:bbb}
\end{equation}
which can be evaluated from experiment.  Actual data on ${\rm Im}F(s)$ are available
only from the $\pi\pi$ threshold up to the middle of the resonance region so
other methods must be found in order to generate the high energy component required 
in order to perform the dispersive integral.  For very high energies we can use the
prediction of perturbative QCD
\begin{equation}
F_1(s)\longrightarrow-{64\pi^2\over 9}{F_\pi^2\over s\ln{-s\over \Lambda^2}}
\end{equation} 
which corresponds to the asymptotic form
\begin{equation}
{\rm Im}F_1(s)\longrightarrow-{64\pi^3\over 9}{F_\pi^2
\over s\ln^2{s\over \Lambda^2}}\label{eq:mmm}
\end{equation}

\begin{figure}
\begin{center}
\epsfig{file=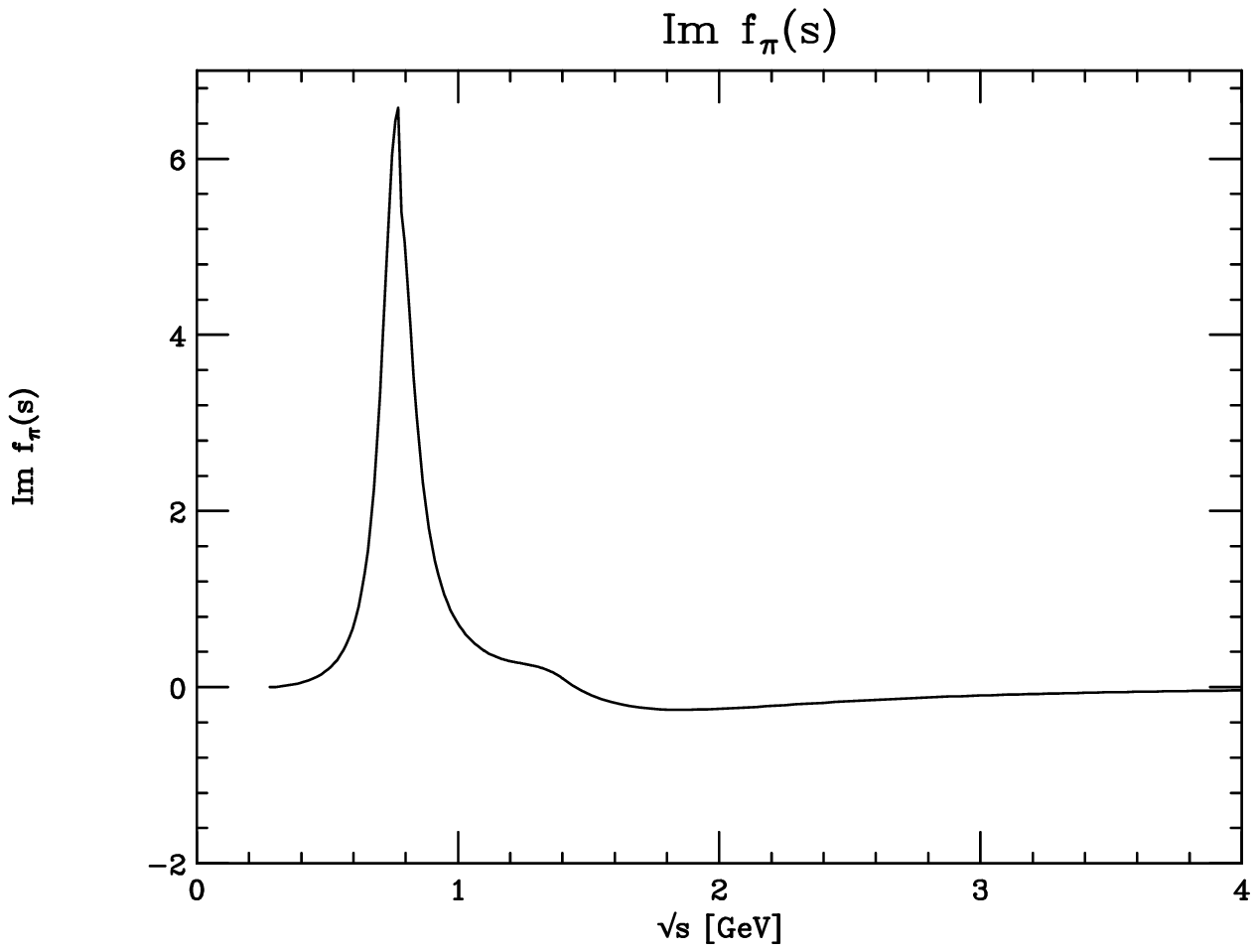,height=6cm,width=10cm}
\caption{Input data for the pion form factor dispersion relation.}
\end{center}
\end{figure}

Knowing the low and high energy forms of the function, we can generate a smooth 
matching which joins them in the intermediate energy region, as shown
in Figure 10\cite{dok}.
Obviously the most striking part of this function is the strong $\rho$ peak, but
note also the long negative tail at high energy.  That the high energy component
must be negative is easy to see from the fact that Eq. \ref{eq:mmm} guarantees 
that the dispersion integral converges even without any subtractions.  Then, taking
the $q^2\rightarrow\infty$ limit produces two additional sum rules   
\begin{eqnarray}
1&=&{1\over \pi}\int_{4m_\pi^2}^\infty{ds\over s}{\rm Im F_1(s)}\nonumber\\
0&=&{1\over \pi}\int_{4m_\pi^2}^\infty ds{\rm Im}F_1(s)\label{eq:ccc}
\end{eqnarray}
In particular the latter sum rule requires that the large positive contribution generated
by integrating over the rho peak be cancelled by a significant negative result from
higher energy.  Now in fact both sum rules are satisfied by the spectral function shown
in Figure 10, as they were used in its construction and are responsible for
the small bump to the right of the rho tail.  Having generated a consistent form for
Im $F(s)$ we can now use Eq. \ref{eq:bbb} in order to generate a prediction for
$L_9^r$.  We find
\begin{equation}
L_9^r(m_\eta)|^{\rm disp. rel.}=0.0074\qquad\mbox{compared to}\qquad 
L_9^r(m_\eta)|^{\rm expt.}=0.0071(3)
\end{equation}
obtained from analysis of low energy phenomenology 
(specifically the pion charge radius).  Obviously there is good agreement here, 
confirming the validity of the dispersive approach (and therefore causality!)

However, while we have gained a new understanding of the chiral perturbative results we
have not yet succeeded in extending their validity to higher energy.  Indeed, Figure 6
shows that, while the form factor $F(q^2)$ matches onto the experimental result at
low $q^2$, it is strongly at variance by the time $q^2\approx 400$ MeV$^2$ due to the
stron rise of the rho.  In fact the prominance of the rho peak suggests an oft-used 
approximation---replacement of the spectral density by a simple delta function
\begin{equation}
{\rm Im}F_1(s)=\pi m_\rho^2\delta(s-m_\rho^2)
\end{equation}      
This drastic approximation---called "vector dominance"---does not 
satisfy the barely convergent sum rule Eq.
\ref{eq:ccc}a but explicitly satisfies the better damped Eq. \ref{eq:ccc}b.  With
respect to that for $L_9$ we find
\begin{equation}
<r_\pi^2>=6m_\rho^2\simeq 0.40\,\,{\rm fm}^2
\end{equation}     
in reasonable agreement with experiment.

An approach which does allow us to move to higher energy is the "inverse amplitude 
method," wherein one looks not at the form factor but rather at its 
inverse.\cite{tru},\cite{iva}  
The unitarity condition in this case reads
\begin{equation}
{\rm Im}{1\over F_1(s)}=-{e^{i\delta_1^1(s)}\sin\delta_1^1(s)\over F_1(s)}
=-\sqrt{s-4m_\pi^2\over s}
{T^1_1(s)\over F_1(s)}
\end{equation}
and a doubly subtracted dispersion relation reads
\begin{equation}
{1\over F_1(q^2)}=1-{<r_\pi^2>\over 6}q^2-{1\over \pi}\int_{4m_\pi^2}^\infty 
{ds\over s^2(s-q^2-i\epsilon)}\sqrt{s-4m_\pi^2\over s}{T^1_1(s)\over F_1(s)}
\end{equation}
Using the leading chiral forms, the integration can be performed using Eq. \ref{eq:uu}  
and yields
\begin{equation}
F_1(q^2)={1\over 1-{<r_\pi^2>\over 6}q^2-{1\over 96\pi^2F_\pi^2}\left((q^2-4m_\pi^2)H(q^2)
+{2\over 3}q^2\right)}
\end{equation}
Expanding in $q^2$ we observe that this form is consistent with the ${\cal O}(p^4)$
chiral result and represents summation of the Lippman-Scwinger equation in
terms of a geometric series.  (Sometimes this is also called the Pade (1,1) 
approximant form.\cite{tru})  Phenomenologically it has the right stuff.
Indeed, as shown in Figure 6, the rho resonance arises very naturally as a pole 
as $s\rightarrow m_\rho^2$ and the phase shift, given by $\delta^1_1={\rm arg}F(s)$ 
is in reasonable agreement with experiment.  We stress that it is {\it not} a simple
result of chiral symmetry---rather it is a chirally {\it motivated} form which reduces to
the chiral result at low energy.

It is interesting to note that there is an exact solution to the problem of finding
a function whose phase is $\delta^1_1$ along the real axis from $4m_\pi^2<s<\infty$.  
This is the Omnes solution and has the general form\cite{omn}
\begin{equation}
F_1^{\rm Omnes}(q^2)=P(q^2)\exp {q^2\over \pi}\int_{4m_\pi^2}^\infty{ds\over s}
{\delta^1_1(s)\over s-q^2-i\epsilon}
\end{equation}   
Here $P(s)$ is an arbitrary polynomial which for our case is set equal to unity.
Using {\it experimental} $\pi\pi$ phase shifts one can construct the function  
$F_0(q^2)$ and compare with experiment and with the inverse amplitude form.
Agreement with experiment is basically quite good.  Note that the Omnes solution
assumes the validity of the Fermi-Watson theorem along the entire real axis and
hence ignores the substantial inelasticity which arises above $\sim 1.2$ GeV.  Thus
this form should presumably only be trusted up to energies where such inelasticity
takes over.

One can in similar fashion construct a form for the partial wave scattering amplitude
which is unitary
\begin{equation}
{\rm Im}{1\over T_1^1(s)}=-i\sqrt{s-4m_\pi^2\over s}
\end{equation}  
and agrees with the $\pi\pi$ phase shifts found from the form factor as
\begin{equation}
T_1^1(s)={1\over 96\pi^2F_\pi^2}{(s-4m_\pi^2)\over 1-{<r_\pi^2>\over 6}s-
{1\over 96\pi^2F_\pi^2}\left((s-4m_\pi^2)H(s)+{2\over 3}s\right)}\label{eq:ghi}
\end{equation}
In fact, such a form was written down long before the development of chiral perturbative 
techniques as a simple unitary generalization of the lowest order (Weinberg) scattering
amplitude.\cite{brg}  It is sometimes called the N/D form, 
where D is the inverse form factor and
satisfies the Fermi-Watson theorm along the right-hand cut while N is the lowest
order chiral amplitude and is analytic along this discontinuity.  Of course, 
Eq. \ref{eq:ghi} violates crossing symmetry in that it does not have the 
proper discontinuity along the
left-hand (u-channel) cut, but we may hope that this does not matter along the physical
(right-hand) cut as it is far away.

We see then that by the use of chiral-based methods, one {\it can} 
construct analytic forms
for observables which satisfy the strictures of unitarity exactly (not perturbatively
as in the case of chiral perturbation theory itself) and which reduce to the forms 
demanded by chiral symmetry at low energy.  They are admittedly no longer 
model-independent, nor do they satisfy all the strictures of field theory (such as 
crossing symmetry).  However, any reasonable model must assume a similar form
and they can be used with some confidence to construct a successful phenomenology.

\section{Closing Comments}      
In the preceding lectures we have covered a lot of ground---effective field theory,
basic chiral models, chiral perturbation theory in the meson and baryon sectors, and
high energy extensions based upon dispersion theory.  Nevertheless the discussion has
essentially been at an introductory level and there is lots more to learn for those 
who are interested.  The state of the art both for experiment and for theory can be
found in the proceedings of the first two chiral dynamics workshops (which took 
place at MIT and Mainz
respectively) which were published by Springer-Verlag.\cite{spr}  
The next such meeting, by the way,
will be held this summer at JLab, and you are certainly invited to attend.

Despite all the work which has taken place there is still much to do for those who
wish to take up the challenge.  This lies in many areas:
\begin{itemize}
\item[i)] calculating to yet higher order (two loop) in both the meson and 
baryon sectors;

\item[ii)] making the connection with fundamental theory by calculating the chiral
coefficients directly from QCD by lattice or other methods;

\item[iii)] development of reasonably model-independent methods to extend chiral
results to higher energy;

\item[iv)] identifying and performing the experiments which most sensitively probe such
theories.
\end{itemize}

My basic message, which I hope you take with you, is that chiral perturbative 
methods are an important and vital component of contemporary particle/nuclear
physics and that at present they allow perhaps the best way to confront low energy
phenomenology with the fundamental QCD Lagrangian which
presumably underlies it.

\begin{center}
{\bf Acknowledgement}
\end{center}       

It is a pleasure to acknowledge warm hospitality of the organizers of
this meeting.  This work was supported in part by the National Science Foundation.

\end{document}